\documentclass[10pt,journal]{IEEEtran}
\usepackage{amsmath,amsfonts}
\usepackage{array}

\usepackage{textcomp}
\usepackage{stfloats}
\usepackage{url}
\usepackage{verbatim}
\usepackage{graphicx}
\usepackage{cite}
\usepackage[numbers,sort&compress]{natbib} 

\usepackage{xcolor}
\usepackage[pdfpagemode=UseNone,
	bookmarksopen=false,
	colorlinks=true,
	urlcolor=blue,
	citecolor=magenta,
	citebordercolor=blue,
	linkcolor=blue, 
	urlcolor=cyan, 
	filecolor=red]{hyperref}
\usepackage{amssymb}
\usepackage{graphicx} 
\usepackage{bm} 
\usepackage{multirow} 
\usepackage{subfigure}
\usepackage{makecell}

\makeatletter

\newcommand{\Rmnum}[1]{\expandafter\@slowromancap\romannumeral #1@}
\makeatother

\usepackage{amsthm}

\newtheorem{proposition}{Proposition}

\newtheorem{remark}{Remark}
\newtheorem{example}{Example}

\usepackage[ruled,linesnumbered]{algorithm2e}
\makeatletter
\newcommand{\removelatexerror}{\let\@latex@error\@gobble}
\makeatother

\begin{document}
%
\title{Joint Transmission for Cellular Networks with Pinching Antennas: System Design and Analysis}

\author{Enzhi~Zhou, ~\IEEEmembership{Member,~IEEE}, Jingjing~Cui, ~\IEEEmembership{Senior Member,~IEEE} and Ziyue~Liu, \\ Zhiguo Ding,  ~\IEEEmembership{Fellow,~IEEE} and Pingzhi Fan,  ~\IEEEmembership{Fellow,~IEEE}
\vspace{-2.0em}
	\thanks{E. Zhou and Z Liu are with the Department
		of Computer and Software Engineering, Xihua University, Sichuan, China (e-mail: ezzhou@mail.xhu.edu.cn, liuziyue\_2006@126.com).}
	\thanks{J. Cui and P. Fan are with the School of Information Science and Technology,
Southwest Jiaotong University, Chengdu 610031, China (e-mail:
jingjing.cui@swjtu.edu.cn; p.fan@ieee.org).}
	\thanks{Z. Ding is with the Department of Computer and Information Engineering, Khalifa University, Abu Dhabi, UAE, and the Department of Electrical and Electronic Engineering, the University of Manchester, Manchester, UK. (e-mail:zhiguo.ding@manchester.ac.uk).}

}

\maketitle

\vspace{-4.0em}
\begin{abstract}
 As an emerging flexible antenna technology for wireless communications, pinching-antenna systems, offer distinct advantages in terms of cost efficiency and deployment flexibility. This paper investigates joint transmission strategies of the base station (BS) and pinching antennas (PAS), focusing specifically on how to cooperate efficiently between the BS and waveguide-mounted pinching antennas for enhancing the performance of the user equipment (UE). By jointly considering the performance, flexibility, and complexity, we propose three  joint BS-PAS transmission schemes along with the best beamforming designs, namely standalone deployment (SD), semi-cooperative deployment (SCD) and full-cooperative deployment (FCD). More specifically, for each BS-PAS joint transmission scheme, we conduct a comprehensive performance analysis in terms of the power allocation strategy, beamforming design, and practical implementation considerations. We also derive closed-form expressions for the average received SNR across the proposed BS-PAS joint transmission schemes, which are verified through Monte Carlo simulations. Finally, numerical results demonstrate that deploying pinching antennas in cellular networks, particularly through cooperation between the BS and PAS, can achieve significant performance gains. We further identify and characterize the key network parameters that influence the performance, providing insights for deploying pinching antennas.

\end{abstract}

\begin{IEEEkeywords}
Pinching antennas (PAS), BS-PAS joint transmission schemes, performance analysis
\end{IEEEkeywords}

\IEEEpeerreviewmaketitle

\vspace{-0.5em}
\section{Introduction}

The advent of the sixth generation (6G) networks opens up a new venue of revolutionary advancements in mobile communications, for improving communication efficiency and expanding coverage in pervasive communication areas. With this momentum,  multiple-input and multiple-output (MIMO) \cite{foschini1996layered} and massive MIMO \cite{larsson2014massive} techniques were deployed to provide high multiplexing gains and space division multiple access, which lays fundamental principles of providing high-speed  and large-connectivity transmissions.  In order to provide more flexibility for antenna configuration,  emerging flexible-antenna techniques, such as reconfigurable intelligent surfaces \cite{di2020smart}, dynamic metasurface antennas \cite{shlezinger2021dynamic}, fluid antennas \cite{new2024tutorial}, movable antennas \cite{zhu2023movable}, pinching antennas (PAS) \cite{suzuki2022pinching} and among others, have attracted widespread interest from academia and industry. Compared to their peers, the PAS possess a number of good features, including flexible deployment that enables rapid establishment of mobile communication coverage, low-cost configuration that can be created by applying simple dielectric particles,  e.g., clothes pinches on  waveguides, and  thus  pinching antennas have become a promising  technique to create strong line-of-sight (LoS) connections quickly and economically \cite{suzuki2022pinching,yang2025pinching,ding2025blockage}.

When the line-of-sight (LoS) path between base stations (BS) and user equipment (UE) is obstructed, the use of PAS can create alternative LoS channels through waveguide-extended deployment and low-cost PAS, thereby enhancing received signal power at UEs. To this end, this paper focuses on how to design the joint transmission scheme by leveraging both the massive antenna arrays at the BS and the flexible deployment advantages of PAS to achieve superior performance gains.

 Motivated by the demonstration from DOCOMO 2022 \cite{suzuki2022pinching}, PAS with the features of fast and cost-effective construction has been viewed as one of propitious flexible antenna
technologies to provide seamless coverage that has attracted great research efforts \cite{yang2025pinching} and \cite{liu2025pinching}.  The basic design principles for  pinching-antenna  systems were proposed in \cite{ding25flexible}  from the perspective of practical deployment, in which  analytical performance comparisons were conducted by using both conventional orthogonal multiple access and non-orthogonal multiple access (NOMA), respectively.  Following this principle,  the pinching-antenna locations were optimized by maximizing the sum rate of the system \cite{xu2025rate}, where multiple PAS were deployed on a single waveguide to serve a single user. The array gain of pinching-antenna systems  was investigated in \cite{ouyang2025array} with respect to the number of PAS and the inter-antenna spacing. 
As a further advance, NOMA assisted pinching-antenna systems with multiple waveguides and each having a single pinching antenna in \cite{hu2025sum}.  

For uplink pinching-antenna systems,  the sum rate maximization problem also have attracted a lot  of research interest in  different  scenarios  \cite{zeng2025sum,zhang2025uplink,bereyhi2025mimo}. More explicitly, the  sum rate was optimized concerning the position of PAS and the power allocation of the users for  a pinching-antenna assisted  NOMA system with one waveguide and one pinching antenna.   \cite{zhang2025uplink} considered an uplink pinching-antenna assisted  multiuser multiple input single-output (MISO) system with multiple waveguides and each having one pinching antenna, in which  the sum rate was maximized by designing the power of the users and the positions of the PAS on each waveguide. 
In   \cite{bereyhi2025mimo},  a pinching-antenna system with multiuser multiple-input multiple-output (MIMO) was considered, where the digital precoding matrix and the location of the PAS on each waveguide were jointly designed to maximize a weighted sum rate. 

Due to their superior comparability, PAS can be applied to various communication systems. For example,  \cite{badarneh2025physical} considered the physical-layer security system with PAS, which demonstrates that the secrecy capacity between the  base station and the legitimate user  increases when the height of the pinching-antenna is closer to the user. Furthermore,  the application of  PAS to covert  communications was explored in \cite{jiang2025pinching}. Another  appealing direction is to explore the benefits of how to use PAS to support integrated sensing and communications \cite{ding2025pinching,bozanis2025cramer} in terms of the Cram\'er–Rao lower bound (CRLB).  Considering the low-cost and flexible reconfigurability features of PAS,  \cite{ding2025pinching} firstly derived the CRLB of the pinching-antenna assisted integrated sensing and
communication (ISAC) systems, revealing the potential benefits in realizing flexible user-centric positioning. More sophisticated research work has been recently conducted in \cite{bozanis2025cramer,khalili2025pinching,zhang2025integrated} for further exploring the benefits of PAS in ISAC systems.

We note that PAS can be treated as a distributed-antenna system  \cite{Heath13a}  with PAS distributed on spatially separated waveguides that are connected to the BS via an interface unit. Distributed-antenna systems were originally proposed to cover the coverage dead spots in indoor communications \cite{Saleh87distributed},  
and design specifications of distributed-antenna systems were further provided in \cite{DASWG23design}. 
Therefore, it is possible that the joint transmission scheme can leverage both the massive antenna arrays at BS and the flexible deployment advantages of PAS to achieve superior performance gains. 
Similar to cooperative reconfigurable intelligent surface (RIS) systems, where the phase shifts of RIS elements were designed to manipulate reflected signals \cite{Wang21Joint}, the BS-PAS system  achieves efficient joint transmission by appropriately designing the phases of  signals  transmitted from the PAS. However, cooperative transmission between the BS and the PAS typically requires the exchange of channel state information (CSI) and joint baseband processing  \cite{Yi24Distributed}.
Since the multi-antenna BS and the PAS operate  independently at the baseband,  enabling joint signal processing would inevitably lead to increased system complexity. 
Consequently, the deployment of the PAS in existing cellular networks should minimize modifications to current baseband and radio frequency (RF) systems. 
To this end, this paper investigates BS-PAS joint transmission architectures, analyzes their feasibility and trade-offs between the performance and complexity. We propose beamforming designs tailored to various BS-PAS deployment configurations considered in this paper, and conduct a comprehensive performance analysis,   providing practical guidelines for integrating PAS into cellular networks. The main contributions of this paper are as follows.
\begin{itemize}

    \item We propose three BS-PAS joint transmission schemes in terms of the coordination levels of signal processing  between the BS and PAS. Based on considerations of communication performance, deployment flexibility, and network complexity, we propose three distinct joint BS-PAS transmission architectures: standalone deployment (SD), semi-cooperative deployment (SCD) and full-cooperative deployment (FCD).
	
	\item  We design the optimal beamforming vectors based on the maximum ratio transmission (MRT) principle, tailored to each of the three proposed joint transmission schemes. 
    In particular, we analyze how the design of pinching-antenna positions facilitates the derivation of the corresponding beamforming vectors.  
    As coordination among the antennas increases across the SD, SCD and FCD schemes, the system performance improves  at the cost of higher implementation complexity.

	\item We derive closed-form expressions of the average received signal-to-noise ratio (SNR) based on  the designed optimal beamforming vectors for each of the three proposed joint BS-PAS transmission schemes. We then analyze the performance gains of three schemes compared to the conventional BS-only approach, providing mathematical expressions in closed-form that quantify the achievable improvements.

    \item Monte Carlo simulations are conducted to validate  the developed results on the average received SNR for the three proposed BS-PAS joint transmission schemes. Our numerical evaluations demonstrate that  the three proposed BS-PAS joint transmission schemes achieve substantial performance improvements in most scenarios compared to the conventional BS-only scheme.

\end{itemize}

\section{System Model and Channel Characterization}
\label{sec:sysmod_channel}

\subsection{System model}
\label{sec:sysmod}
\begin{figure}
	\vspace{-0.1cm}
	\centering
	\includegraphics[width=0.4\textwidth]{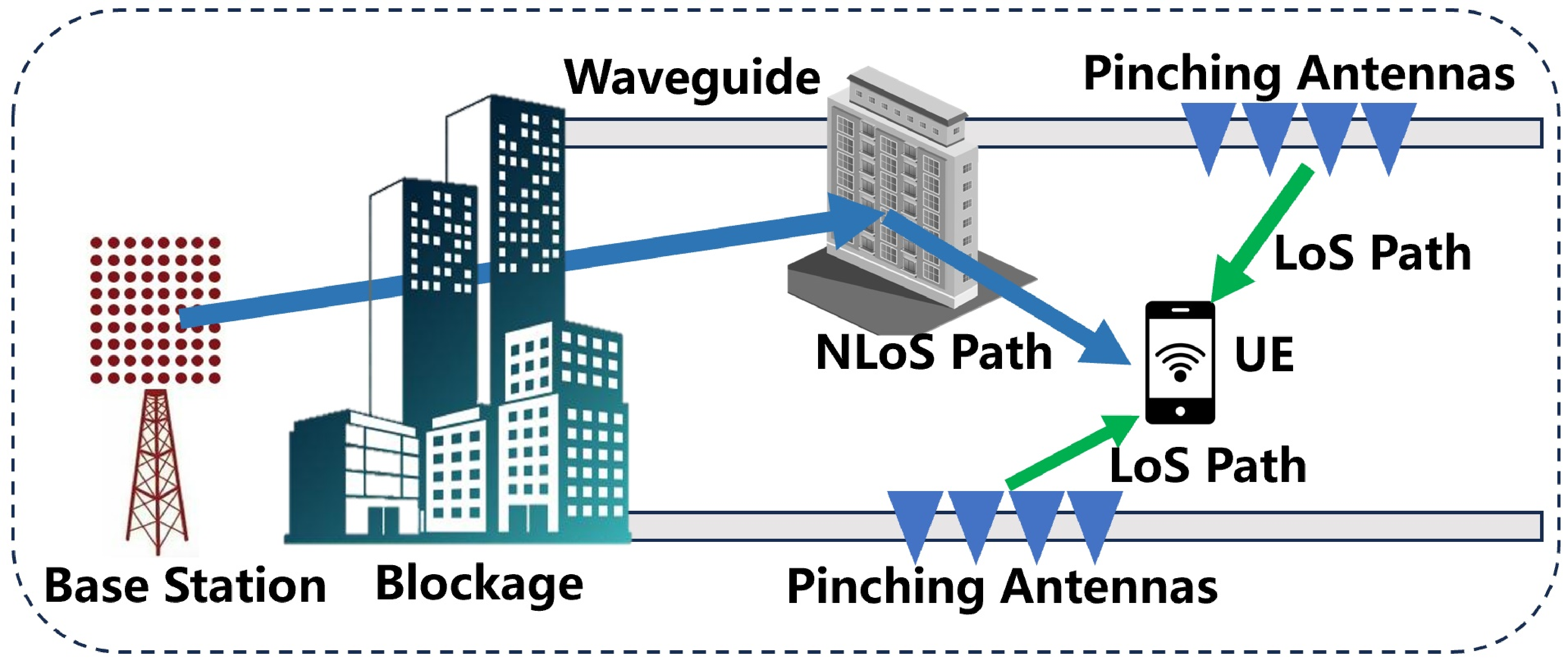}
	\caption{The joint transmission scenario of the BS and PAS.}\label{fig:system_model}
       \vspace{-1.5em}
\end{figure}

In this paper, we consider a downlink pinching-antenna assisted communication scenario as shown in Fig. \ref{fig:system_model}, where the BS equipped with $N_B$ antennas communicates with one single-antenna users with the assistance of PAS. There are $K$ waveguides deployed,  with $N_G$ PAS activated on each waveguide.  Assume that the LoS path between the BS and the UE is blocked by some obstacles. As a result,  the user equipment receives  the transmitted signals from the BS via NLoS propagation and from the PAS via LoS paths, which requires the cooperative transmission by the BS and PAS.  

\subsection{Channel characterization}

Following \cite{suzuki2022pinching}, the channel between a waveguide-mounted pinching antenna and the UE can typically be characterized as a LoS channel.  To evaluate the impact of different propagation environments between the BS-UE and PAS-UE channels, we employ distinct path loss exponents to capture these different channel characteristics \cite{rappaport2024wireless}. 

\paragraph{BS-UE channel model}
The channel between the BS and the UE can be represented as 
\begin{equation} \label{eq:BS_channel}
{{\bm{h}}_B} = \sqrt \frac{{ \eta  }}{L_B^{\alpha }}{\widetilde {\bm{h}}_B},
\end{equation}
where ${\widetilde {\bm{h}}_B} = \left[ {{{\tilde h}_{B,1}}, \ldots ,{{\tilde h}_{B,{N_B}}}} \right]$ and $\eta  = \frac{{{c^2}}}{{16{\pi ^2}f_c^2}}$.  $c$ and $f_c$ are the speed of light and  the carrier frequency, respectively.  $L_B$ is the distance between BS and the UE and $\alpha$ is the path loss exponent \cite{chaudhari2020lpwan}, characterizing the rate at which signal power attenuates with propagation distance.
 Moreover, ${\tilde h_{B,i}} \in \mathbb{C}$ is the channel coefficient between the $i$-th BS antenna and the UE, which is modeled as the Rayleigh fading channel, where the amplitude of channel gain $\left| {\tilde h_{B,i}} \right|$ follows a Rayleigh distribution with scale parameter $\sigma^2 = \frac{1}{2}$, i.e., $ \left|{\tilde h_{B,i}} \right| \sim \mathrm{Rayleigh}\left(\frac{1}{\sqrt{2}}\right) $.

\paragraph{PAS-UE channel model}
Let ${\bm{\varphi }}_{f,k}$ denote the coordinate of the signal injection point for waveguide $k$, while ${\bm{\varphi }}_{k,n}$ represents the  coordinate of the $n$-th pinching antenna on waveguide $k$. Assume that the UE is located at position ${\bm{\varphi }}_u$. The phase delay experienced by the signal injected into waveguide $k$ when propagating through the $n$-th pinching antenna to reach the UE can be expressed as 
\begin{equation}
{\psi_{k,n}} =  {\frac{{2\pi }}{\lambda }\left\| {{{\bm{\varphi }}_{k,n}} - {\bm{u}}} \right\| + \frac{{2\pi }}{{{\lambda _G}}}\left\| {{{\bm{\varphi }}_{k,n}} - {{\bm{\varphi }}_{f,k}}} \right\|},
\end{equation}
where $\lambda = \frac{2\pi}{f_c}$ is the wavelength in the free space, ${\lambda _G} = \frac{\lambda }{{{n_{eff}}}}$ represents the wavelength of the signal within the waveguide,  and $n_{eff}$ denotes the effective refractive index of the dielectric waveguide. 

Following the results in \cite{xu2025rate}, the  positions for PAS are assumed to be consistently clustered at the location closest to the UE along the waveguide, requiring only wavelength-level fine tuning of the antenna positions to achieve the maximum antenna gain. Since the signal wavelength in the band of wireless cellular networks is typically less than $100$ millimeters, such positional adjustments has a negligible affect on the large-scale path loss. Additionally, we neglect the transmission loss inside the waveguide as in \cite{liu2025pinching}.
Thus,  the PAS-UE channel through the $k$-th waveguide is given by 
\begin{equation}
\begin{aligned}
{h_{G,k}} &= \frac{{\sqrt \eta  }}{{\left\| {{{\bm{\varphi }}_{k,n}} - {\bm{u}}} \right\|}} \times \sqrt {\frac{1}{{{N_G}}}} \sum\limits_{n = 1}^{{N_G}} {{e^{ - j{\psi _{k,n}}}}} \\
&= \sqrt\frac{{ \eta  }}{{{L_{G,k}^\beta}{{N_G}} }}\sum\limits_{n = 1}^{{N_G}} {{e^{ - j{\psi_{k,n}}}}},
\end{aligned}
\end{equation}
 where $L_{G,k}$  is the distance between PAS on the $k$-th waveguide and the UE, and $\beta$ is the associated path loss exponent.  Thus, $L_{G,k}^\beta$ represents the path loss between PAS on the $k$-th waveguide and the UE. Moreover, we consider that 
each antenna at the BS is equipped with a dedicated RF chain, whereas each waveguide -- despite containing multiple PAS -- is served by a single  RF chain. Consequently, the joint downlink channel of the UE from the BS and PAS  can be modeled as a MISO channel with a total of $N_B + K$ transmit antennas.

\section{Proposed BS-PAS Joint Transmission Schemes }

In this section, we start by introducing the three proposed joint transmission schemes, categorized according to the  level of cooperation between the BS and PAS. This is followed by an analysis of the optimal deployment of PAS on each waveguides, which in turn motivates the use of  a joint transmission channel model for the BS-PAS cooperative system, serving as the foundation for designing optimal beamforming vectors.

\subsection{Proposed joint transmission schemes}

\begin{figure}[h]
	\centering
	\hspace{-5pt}
	\subfigure[Standalone Deployment.]{
		\label{fig:standalone_deployment}
		\includegraphics[width=0.4\textwidth]{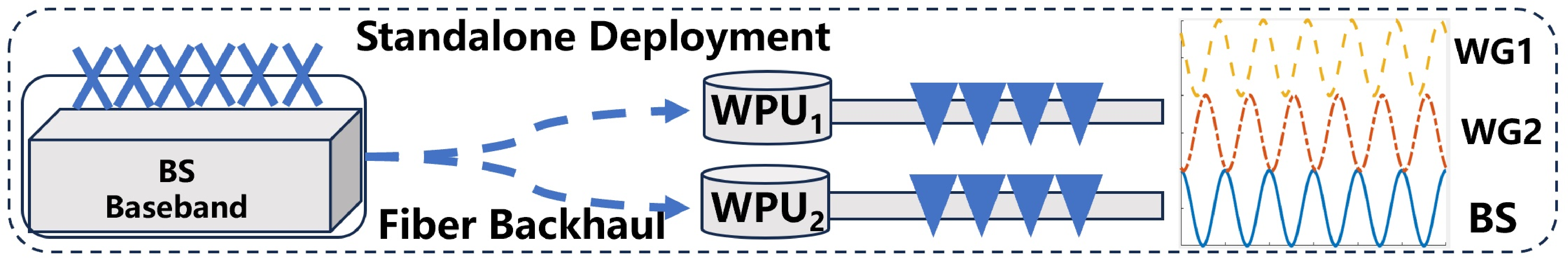}
	}\hspace{-5pt}
	\subfigure[Semi-cooperative Deployment.]{
	\label{fig:semi-cooperative deployment}
	\includegraphics[width=0.4\textwidth]{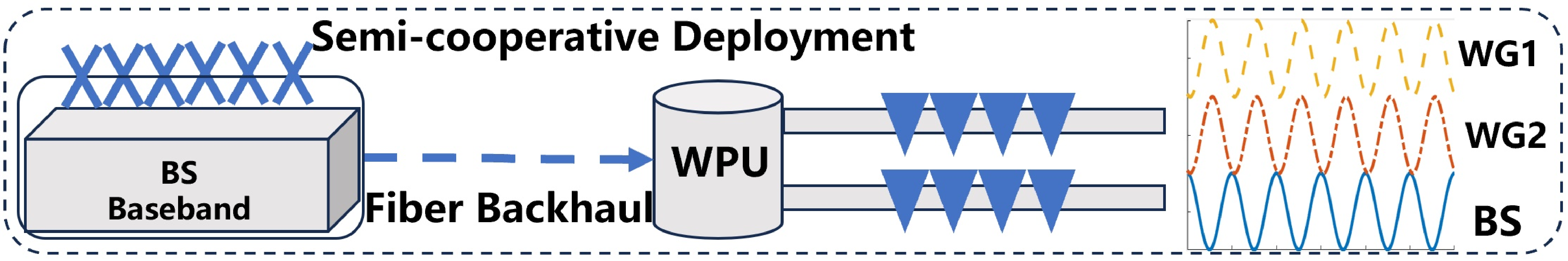}
	}\hspace{-5pt}
	\subfigure[Full-cooperative Deployment.]{
	\label{fig:joint_deployment}
	\includegraphics[width=0.4\textwidth]{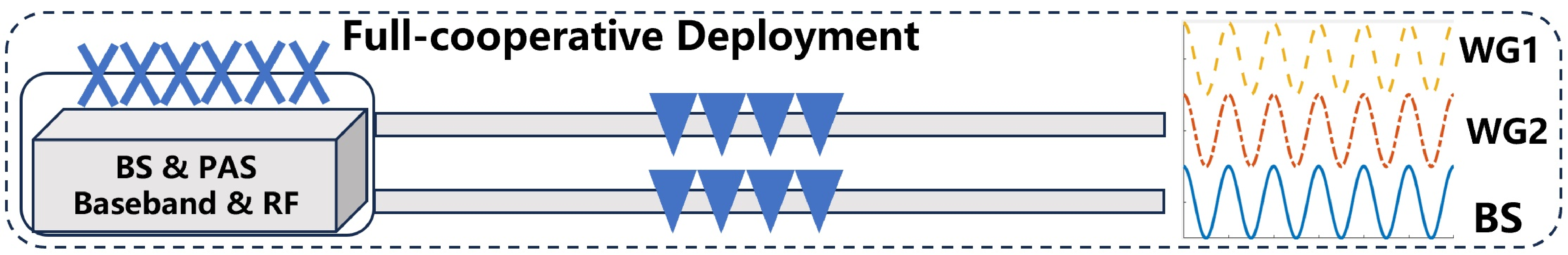}
	}
	\caption{The architectures of BS-PAS joint transmission system.}\label{fig:architecture}
    \vspace{-0.7em}
	\end{figure}

Inspired by the specification of distributed-antenna system \cite{DASWG23design}, we treat pinching-antenna systems with multiple waveguides as a cooperative infrastructure in this paper, which can be operated in three different manners: 1)  the BS signal  distributes the signal to separate waveguide processing units (WPUs) that can be deployed in different locations, in which each WPU independently feeds the signal to PAS on each waveguide (e.g., optical fiber cables or dedicated structured cabling) for improving the communication performance and coverage of the system; 2) the signal transmit from the BS is connected to a common WPU that is deployed in some place different from the BS, which then distributed the signal to each waveguide simultaneously; 3) the WPU is integrated in the BS, in which the BS signal is fed to the waveguides directly from the BS without further intermediate interface.
Correspondingly,  three BS-PAS joint transmission deployment architectures are developed: standalone deployment, semi-cooperative deployment, and joint deployment. The three joint transmission architectures are illustrated in Fig. \ref{fig:architecture}. 

\paragraph{Standalone deployment (SD)}
In the SD scheme, as shown in Fig. \ref{fig:standalone_deployment} with two waveguides: WG1 and WG2, it places separate waveguide processing units (WPUs) in different service areas, connected to the BS via optical fibers. This architecture offers maximum deployment flexibility of PAS while reducing power loss and interference risks from long-distance waveguide transmission. In this scheme, each WPU needs to obtain UE location information and baseband signals from the BS, then adjusts its pinching-antenna positions to generate and inject radio-frequency signals into the waveguide. 
 Each waveguide independently adjusts its PAS, resulting in phase differences of the received signals from different waveguides and BS, as shown in  Fig. \ref{fig:standalone_deployment}.

\paragraph{Semi-cooperative deployment (SCD)} 
As shown in Fig. \ref{fig:semi-cooperative deployment}, in the SCD scheme, the processing units of all waveguides are co-located, allowing for joint signal processing among all PAS. Here, the centralized WPU enables power allocation among different waveguides. Additionally, multiple waveguides can coordinately adjust pinching-antenna positions to ensure in-phase signal superposition at the receiver for the maximum array gain of PAS.  However, this enhanced coordination comes at the cost of reduced PAS deployment flexibility. Compared to the SD scheme, the joint WPU in SCD enables joint adjustments of PAS to  achieve signal-phase alignment across all waveguides, resulting in increased WPU complexity.

\paragraph{Full-cooperative deployment (FCD)}
In the FCD scheme, the WPUs are incorporated into the BS, facilitating  fully joint signal processing across the BS and all PAS. This enables flexible power allocation between the BS antennas and waveguides, as well as coordinated placement of PAS, thereby achieving maximum beamforming gain.  Nevertheless, the FCD scheme requires significant modifications to existing BS baseband processing, and waveguide routing from the BS reduces the flexibility of PAS while increasing deployment costs. Moreover, full cooperation between the BS and waveguides introduces additional challenges in the placement and configuration of PAS. Due to the non-idealities of oscillators and antennas, the transmitted signal typically experiences phase noise, which can be estimated at the receiver using demodulation reference signals (DMRS) and subsequently pre-compensated at the transmitter  \cite{Corvaja16Phase, Zhang25Evaluating}. This implies that in order to ensure phase alignment with the BS-transmitted signal, the PAS must dynamically adjust their positions to match the BS signal phase. This requirement introduces additional signal processing complexity and poses potential risk of delayed adjustments for the PAS. 

We summaries the features of three joint transmission architectures in Table. \ref{table:architecture_characteristic}.  The coordination levels and implementation complexity are progressively increasing across the three architectures. Therefore, practical network deployment should seek a balance between performance requirements and implementation complexity by selecting the most appropriate architecture.

\begin{table}[h]
    \vspace*{-0.7em}
	\centering
	\caption{Characteristics of different joint transmission architectures}\label{table:architecture_characteristic}
	\begin{tabular}{|c|c|c|c|}
		\hline
		\makecell{Architectures}& \makecell{Deployment \\ flexibility} & \makecell{Implementation \\ complexity} & \makecell{Signal coordination \\ level}\\
		\hline
		SD & High & Low & Non-coherent  \\
		\hline
		SCD & Medium & Medium & Partial coherent \\
		\hline
		FCD & Low & High & Ideal coherent \\
		\hline
	\end{tabular}
    \vspace*{-0.7em}
\end{table}

\vspace*{-0.7 em}
\subsection{The joint BS-PAS channel}

Let $\varphi_{k,1}$ and $\psi_{k,1}$  denote the position and the induced phase of the reference pinching antenna on the $k$-th waveguide, respectively. This reference pinching antenna serves as the phase alignment anchor for determining the position of the remaining PAS on the same waveguide. More specifically, the positions of the remaining PAS can  be identified to satisfy the phase coherence condition, thereby ensuring constructive signal combination at the UE \cite{xu2025rate,lv2025beam}. Specifically, for $n=2,\ldots,N_G$, there exists an integer $\tilde k_n$ such that the following condition holds.
{\small
\begin{equation}\label{eq:waveguide_phase_align}
\begin{aligned}
&\frac{{2\pi }}{\lambda }\left\| {{{\bm{\varphi }}_{k,n}} - {\bm{u}}} \right\| + \frac{{2\pi }}{{{\lambda _G}}} \left\| {{{\bm{\varphi }}_{k,n}} - {{\bm{\varphi }}_{f,k}}} \right\| \\
=& \frac{{2\pi }}{\lambda }\left\| {{{\bm{\varphi }}_{k,1}} - {\bm{u}}} \right\| + \frac{{2\pi }}{{{\lambda _G}}} \left\| {{{\bm{\varphi }}_{k,1}} - {{\bm{\varphi }}_{f,k}}} \right\| + 2{\tilde k_n}\pi, \tilde k_n \in Z. \\
\end{aligned}
\end{equation}
}

In this case, the waveguide can be abstracted as an RF port, and the equivalent channel from waveguide RF port $k$ to the UE can be expressed as
\begin{equation}\label{eq:waveguide_channel}
{h_{G,k}} = \sqrt\frac{{ \eta  }}{{L_{G,k}^{{\beta}} {{N_G}} }}\sum\limits_{n = 1}^{{N_G}} {{e^{ - j\left( {{\psi _{k,1}} + 2{k_n}\pi } \right)}}}  = \sqrt\frac{{ {\eta {N_G}} }}{{{L_{G,k}^\beta}}} {{e^{ - j{\phi_k}}}},
\end{equation}
 where ${\phi _k=\psi _{k,1}}$ represents the phase delay of the signal transmitted through the reference antenna of the $k$-th waveguide and arriving at the UE. Next, we give a vanilla example for clarity.

\begin{example} \label{ex:waveguide_phase}
 Consider a dense urban scenario, to ensure coverage for the UE whose LoS channel from the BS is obstructed by buildings, two waveguides are strategically deployed on the facades of opposing buildings.  The UE is assumed to be on the ground at position  $\left( 30, 5, 0\right)$. 
 The feed point positions of the two waveguides are set to ${\bm{\varphi} _{f,1}} = \left( {0,0,10} \right)$ and ${\bm{\varphi} _{f,2}} = \left( {0,30,10} \right)$, respectively. The coordinates of the reference PAS are ${\bm{\varphi}_{1,1}} = \left( {20,0,10} \right)$ and ${\bm{\varphi} _{2,1}} = \left( {20,30,10} \right)$, respectively. Let $f_c = 3.5$ GHz and $n_{eff} = 1.5$.   The corresponding phase anchors of the two waveguides can then be calculated as $\phi_1=2.748$ and $\phi_2=0.846$, respectively.
\end{example}

By combining \eqref{eq:BS_channel} and \eqref{eq:waveguide_channel}, the aggregate channel from both the BS and pinching antennas to the UE can be expressed as
\begin{equation}\label{eq:joint_channel}
{\bm{h}} = \left[ {\sqrt {\frac{\eta }{{L_B^\alpha }}}} {{{\bm{\tilde h}}}_B},\sqrt {\frac{{\eta {N_G}}}{{L_{G,1}^{\beta}}}} {e^{ - j{\phi _1}}},, \ldots ,\sqrt {\frac{{\eta {N_G}}}{{L_{G,K}^{\beta}}}} {e^{ - j{\phi _K}}} \right].
\end{equation}
We will evaluate the performance of beamforming schemes using the average received SNR. The received signal can be expressed as 
\begin{equation}\label{eq:receive_signal_model}
    y = {\mathbf{hw}}s + \omega,
\end{equation}
where $\omega \sim \mathcal{C}\left( 0, \sigma_n^2\right)$ is  white complex Gaussian noise with variance $\sigma_n^2$. $s$ and $\bm{w}$ are the power normalized transmitted symbol and beamforming vector, respectively, i.e., $\left| s\right|^2 = 1 $ and $\left\| \bm{h} \right\|^2=1$. Assuming that the total transmit power is $P_t$, the average received SNR is given by 
\begin{equation}\label{eq:SNR_express}
    \gamma = \frac{ E \left\{ \left| {\mathbf{hw}} \right|^2 \right\}  P_t}{\sigma_n^2},
\end{equation}
where $\left| {\mathbf{hw}} \right|^2$ is the channel gain, which is determined by different BS-PAS architectures and their beamforming schemes.

\section{Proposed Beamforming Schemes and Performance Analysis}
\label{sec:proposed_BF}

In this section, we first investigate the power allocation schemes among RF ports between the BS and waveguides under three BS-PAS joint transmission architectures to facilitate the derivation of beamforming vectors for each scheme. Subsequently, we present the corresponding beamforming designs for these three architectures and derive the expressions in closed-form in terms of equivalent channels and average SNR at the receiver side.

\subsection{Power Allocation Scheme Between the BS and Waveguides}
\label{sec:pow-alloca-BS-waveguides}
In practical implementation, the power allocation flexibility of joint BS-PAS transmission is restricted by the physical deployment and information sharing. In this paper, the power coefficients are used to quantify the proportion of transmit power allocated to  the BS and PAS relative to the total power budget.

As discussed in Section \ref{sec:sysmod_channel}, each BS antenna and each waveguide corresponds to one RF chain. Thus, there are $N_B+K$ RF chains in total in the BS-PAS joint transmission system. To ensure consistency in the total transmit power, in SD and SCD schemes, power allocation across the BS and waveguides is performed based on the proportions of their RF chains, i.e., . ${P_{C,B}} = \frac{{{N_B}}}{{{N_B} + K}}$ and ${P_{C,B}} = \frac{{{K}}}{{{N_B} + K}}$. 
In the SD scheme, the channel condition of each waveguide is not shared, then each waveguide receive an equal fraction of the power given to PAS. In the SCD scheme, all waveguides engage in joint channel processing, necessitating a coordinated power allocation among waveguides according to their CSI. In the FCD scheme, the BS and PAS perform joint signal processing, and have the ability to flexibly allocate power among $N_B + K$ RF chains based on the full CSI from BS antennas/pinching antennas to the UE. 
The power coefficients of joint transmission architectures are shown in Table. \ref{tab:power_coefficients}. 
Following the predefined power allocation strategy as discussed above, we will investigate the optimal beamforming schemes for each transmission strategy, respectively, in the following section.

\begin{table}
	\centering
	\caption{Power coefficients for different joint BS-PAS transmission schemes}
    \label{tab:power_coefficients}
	\begin{tabular}{|c|c|}
		\hline
		Architectures & Power Allocation \\
		\hline
		SD & \makecell{ ${P_{S,B}} = \frac{{{N_B}}}{{{N_B} + K}}$ for BS, \\ ${P_{S,G}} = \frac{1}{{{N_B} + K}}$ for each waveguide } \\
		\hline
		SCD &  \makecell{ ${P_{C,B}} =\frac{{{N_B}}}{{{N_B} + K}}$ for BS, \\ ${P_{C,G}} = \frac{K}{{{N_B} + K}}$ for all waveguides }\\
		\hline
		FCD & \makecell{ Dynamic power allocation \\ among the $N_B + K$ RF chains} \\
		\hline
	\end{tabular}
\end{table}

\subsection{Beamforming design for the SD scheme}

Through uplink channel estimation and leveraging the channel reciprocity between uplink and downlink, the optimal downlink beamforming vector at the BS is maximum ratio transmission (MRT) \cite{Corvaja16Phase}. Incorporating the power coefficients, the optimal beamforming vector at the BS can be expressed as
\begin{equation}
{{\bm{\tilde w}}_B} = \sqrt {{P_{S,B}}} \frac{{{{\bm{h}}_B}^H}}{{\left\| {{{\bm{h}}_B}^H} \right\|}} = \sqrt {\frac{{P_{S,B}}}{{{{\left\| {{{{\bm{\tilde h}}}_B}} \right\|}^2}}}}  {{{{\bm{\tilde h}}}_B}^H},
\end{equation}
where each waveguide can adjust the positions of its own $N_G$ pinching antennas. Since the WPUs of each waveguide are independently deployed, joint power allocation cannot be implemented. Therefore, equal power allocation is adopted among different waveguides. Consequently, under the SD scheme, the beamforming vector of the waveguide transmission only fed the modulated symbols into the waveguide at a given transmit power. Combined with the BS's precoding vector, the power normalized beamforming vector for the SD scheme can be expressed as a $N_B+K$ vector
\begin{equation}
{{\bm{w}}_{S}} = \left[ {\sqrt {\frac{{{P_{S,B}}}}{{{{\left\| {{{{\bm{\tilde h}}}_B}} \right\|}^2}}}} {\bm{\tilde h}}_B^H,\sqrt {{P_{S,G}}} , \ldots ,\sqrt {{P_{S,G}}} } \right].
\end{equation}
In this context, the power allocation among the antennas at the BS is conducted within beamforming, but no power allocation is implemented between the BS and the PAS, or among the waveguides themselves.

Based on the joint channel in \eqref{eq:joint_channel}, the  channel observed at the UE can be expressed as
\begin{equation}
\begin{aligned}
{h_S} &= \bm{h} \bm{w}_S\\
 &= \sqrt \frac{{\eta {P_{S,B}}}}{{L_B^\alpha }} \left\| {{{{\bm{\tilde h}}}_B}} \right\| + \sqrt {\eta {N_G}{P_{S,G}}} \sum\limits_{k = 1}^K {\sqrt\frac{1}{{{L_{G,k}^\beta}}}{e^{ - j{\phi _k}}}}.  \\
\end{aligned}
\end{equation}
In the equation, $\sum\nolimits_{k = 1}^K \sqrt\frac{1}{{{L_{G,k}^\beta}}}{e^{ - j{\phi _k}}} $ is determined by the reference antennas of each waveguide and the UE's position. It can be observed that the equivalent channels from different waveguides may combine  constructively when in-phase, or destructively when out-of-phase, at the receiver. When destructive cancellation occurs, the received waveguide signal power at the UE will decrease. Moreover, if the direction of $\sum\nolimits_{k = 1}^K \sqrt\frac{1}{{{L_{G,k}^\beta}}}{e^{ - j{\phi _k}}} $ approaches the negative real axis, the signals from PAS and BS will interfere destructively with each other.

  \begin{proposition}\label{prop:ave-channel-gain-SD}
  The average channel gain for the SD scheme is given by
\begin{equation}\label{eq:channel_power_Gain_standalone}
E\left\{ {{{\left| {{h_S}} \right|}^2}} \right\} = \frac{{\eta N_B^2}}{{L_B^{\alpha }\left( {{N_B} + K} \right)}} + \frac{{\eta {N_G}}}{{\left( {{N_B} + K} \right)}}\sum\limits_{k = 1}^K {\frac{1}{{L_{G,k}^{\beta}}}},
\end{equation}
if $\phi_k$, the channel phase between the reference pinching antenna and the UE, follows a uniform distribution from $0$ to $2\pi$.
\begin{proof}
    The proof is provided in Appendix \ref{proof-prop:ave-channel-gain-SD}.
\end{proof}
 \end{proposition}

According to \eqref{eq:SNR_express}, the average received SNR of SD is given by  
\begin{equation}\label{eq:snr_SD}
    \gamma_{SD} = \frac{{\eta N_B^2 P_t}}{{L_B^{\alpha }\left( {{N_B} + K} \right) \sigma_n^2}} + \frac{{\eta {N_G} P_t}}{{\left( {{N_B} + K} \right) \sigma_n^2 }}\sum\limits_{k = 1}^K {\frac{1}{{L_{G,k}^{\beta}}}}.
\end{equation}

\subsection{Beamforming design for the SCD scheme}
In the SCD scheme, the WPUs can cooperate with each other.  Thus, the system can jointly optimize the positions of PAS on each waveguide,  ensuring that the signals transmitted from the $K$ waveguide RF ports constructively combine in-phase at the UE. Moreover,  power allocation among the waveguides can also be optimized jointly.
Without loss of generality, the reference antenna position of the first waveguide is assumed to  remain fixed, while the reference antenna positions on the remaining $K-1$ waveguides are adjusted to align their phase delays in \eqref{eq:joint_channel} with that of the first waveguide. Explicitly,  suppose that ${\phi _1}$ denotes the phase delay of the first waveguide's signal. Consequently, the reference pinching-antenna position ${{\bm{\varphi }_{k,1}}},k=2,\ldots,K$ can be tuned to satisfy
\begin{equation}\label{eq:wave_Guide_reference_antenna_condition1}
\begin{split}
&\frac{{2\pi }}{\lambda }\left\| {{{\bm{\varphi }}_{k,1}} - {\bm{u}}} \right\| + \frac{{2\pi }}{{{\lambda _G}}}\left\| {{{\bm{\varphi }}_{k,1}} - {{\bm{\varphi }}_{f,k}}} \right\| \\
=&  {\phi _1}   + 2\tilde k\pi ,~~\tilde k \in Z.
\end{split}
\end{equation}
Invoked by the bidirectional search concept, an adjustment scheme performs leftward and rightward searches centered around the existing reference antenna position to identify the nearest positions satisfying \eqref{eq:wave_Guide_reference_antenna_condition1}, denoted by $\tilde \varphi _{k,1}^L$ and $\tilde \varphi _{k,1}^R$. Then the new waveguide positions are determined by
\begin{equation}
    {\tilde{\bm\varphi} _{k,1}} = \left\{ \begin{gathered}
  \tilde{\bm\varphi} _{k,1}^R\quad \left| {\tilde{\bm\varphi} _{k,1}^L - {{\bm\varphi} _{k,1}}} \right| \geqslant \left| {\tilde{\bm\varphi} _{k,1}^R - {{\bm\varphi} _{k,1}}} \right| \hfill \\
  \tilde{\bm\varphi} _{k,1}^L\quad \left| {{\bm\varphi} \varphi _{k,1}^L - {{\bm\varphi} _{k,1}}} \right| < \left| {\tilde {\bm\varphi} _{k,1}^R - {{\bm\varphi} _{k,1}}} \right|. \hfill \\ 
\end{gathered}  \right.
\end{equation}

Following the vanilla example in \textbf{Example \ref{ex:waveguide_phase}}, we provide an concrete example in \textbf{Example \ref{ex:antena_phase}} for illustrating the optimized positions of the PAS in the proposed system.
\begin{example} \label{ex:antena_phase}
    Referring to the previous example, the search method will adjust the position of the reference antenna for  Waveguide 2 to ${{\tilde{\bm\varphi} }_{2,1}} = \left( {20.017,30,10} \right)$, thus enabling Waveguide 2 to have the same phase delay as Waveguide 1, i.e., in $\phi_1=\phi_2=2.748$. Note that the positions of the remaining antennas in Waveguide $2$ also need to  be adjusted accordingly to satisfy condition \eqref{eq:waveguide_phase_align}.
\end{example}

Furthermore, regarding the power allocation ratios among waveguides, we have the following proposition. 
\begin{proposition}\label{prop:scd-pa-channel}
Given that the reference antenna position on each waveguide satisfies the condition in \eqref{eq:wave_Guide_reference_antenna_condition1}, the maximum equivalent channel gain can be achieved when the power allocation ratio of waveguide $k$ is proportional to its channel gain $\frac{{\eta {N_G}}}{{L_{G,k}^{\beta}}}$.
\end{proposition}
\begin{proof}
The proof is provided in Appendix~\ref{proof-prop:scd-pa-channel}. 
\end{proof}

As cooperation only happens among the waveguides, the MRT transmission for the BS is still the optimal beamforming scheme, and thus the joint beamforming vector in this case can be expressed as
{\small
\begin{equation}
\begin{gathered}
{\bm{w}_{C}} =  \hfill \\
\left[ {\sqrt {\frac{{{P_{C,B}}}}{{{{\left\| {{{{\bm{\tilde h}}}_B}} \right\|}^2}}}} {\bm{\tilde h}}_B^H,\sqrt {\frac{{{P_{C,G}}}}{{\sum\limits_{k = 1}^K {\frac{{\eta {N_G}}}{{L_{G,k}^{\beta}}}} }}} \left[ {\sqrt {\frac{{\eta {N_G}}}{{L_{G,1}^{\beta}}}}, \ldots , \sqrt {\frac{{\eta {N_G}}}{{L_{G,K}^{\beta}}}} {e^{ - j\left( {{\phi _1} - {\phi _K}} \right)}}} \right]} \right]. \hfill \\
\end{gathered} 
\end{equation}
}

It should be pointed out that unlike conventional beamforming methods, the beamforming for PAS is implemented by configuring the phase ${\phi _1} - {\phi _k}$, which is achieved by adjusting the position of reference pinching antenna on the $k$-th waveguide. 

Therefore, in the SCD scheme, the equivalent channel at the receiver side can be expressed as 
\begin{equation}
\begin{gathered}
{h_{C}} = {\bm{h}}{{\bm{w}}_{C}} \hfill \\
= \sqrt \frac{\eta {P_{C,B}}}{{L_B^\alpha }} \left\| {{{{\bm{\tilde h}}}_B}} \right\| + \sqrt {\eta {P_{C,G}}\sum\limits_{k = 1}^K {\frac{{{N_G}}}{{L_{G,k}^{\beta}}}} } {e^{ - j{\phi _1}}} \hfill \\ 
\end{gathered},
\end{equation}
where the equivalent channel gain $\sum\nolimits_{k = 1}^K {\frac{{{N_G}}}{{L_{G,k}^{\beta}}}} $ of the PAS is determined by both the number of PAS in each waveguide and their distances to the UE, while the equivalent channel phase delay ${\phi _1}$ is governed by the distance between the reference antenna of the first waveguide and the UE. It can be observed that the signals transmitted from the PAS and BS may constructively interfere in-phase at the receiver. However, when the direction of ${\phi _1}$ approaches $\pi+2\tilde k \pi$, they will mutually cancel each other.

\begin{proposition}\label{prop:svdscheme}
   In the SCD scheme, the average channel gain for the SCD scheme has a closed expression as
\begin{equation}\label{eq:semi_cooperative_channel_power_Gain}
E\left\{ {{{\left| {{h_{C}}} \right|}^2}} \right\} = \frac{{\eta N_B^2}}{{\left( {{N_B} + K} \right)L_B^{\alpha }}} + \frac{{\eta {N_G}K}}{{\left( {{N_B} + K} \right)}}\sum\limits_{k = 1}^K {\frac{1}{{L_{G,k}^{\beta}}}}.
\end{equation}
\begin{proof}
   See Appendix C.
\end{proof}
\end{proposition}

Based on  the results of {\textbf{Proposition \ref{prop:svdscheme}}} and the SNR expression in  \eqref{eq:SNR_express}, we can obtain the average received SNR of the SCD scheme as follows:
\begin{equation}\label{eq:snr_SCD}
    \gamma_{SCD} = \frac{{\eta N_B^2} P_t}{{\left( {{N_B} + K} \right)L_B^{\alpha } \sigma_n^2}} + \frac{{\eta {N_G}K} P_t}{{\left( {{N_B} + K} \right) \sigma_n^2}}\sum\limits_{k = 1}^K {\frac{1}{{L_{G,k}^{\beta}}}}.
\end{equation}

\subsection{Beamforming design for the FCD scheme}
In the FCD scheme, the BS and all waveguides can perform joint signal processing. In this case, the optimal beamforming vector follows the MRT criterion of the joint channel in \eqref{eq:joint_channel}. For simplicity, we neglect the phase noise of the BS's transmitted signals, and the beamforming vector is given by
\begin{equation}\label{eq:precoder_joint}
\begin{gathered}
{{\bm{w}}_F} = \frac{{{{\bm{h}}^H}}}{{\left\| {\bm{h}} \right\|}} \hfill \\
= \frac{{\left[ {\sqrt {\frac{\eta }{{L_B^{\alpha }}}} {{{\bm{\tilde h}}}_B},\sqrt {\frac{{\eta {N_G}}}{{L_{G,1}^{\beta}}}} {e^{j{-\phi _1}}}, \ldots ,\sqrt {\frac{{\eta {N_G}}}{{L_{G,K}^{\beta}}}} {e^{j{-\phi _K}}}} \right]}}{{\sqrt {\frac{\eta }{{L_B^{\alpha }}}{{\left\| {{{{\bm{\tilde h}}}_B}} \right\|}^2} + \eta {N_G}\sum\limits_{k = 1}^K {\frac{1}{{L_{G,k}^{\beta}}}} } }} \hfill, \\ 
\end{gathered}
\end{equation}
where the phase shift term ${e^{j{\phi _k}}}$ in the $k$-th waveguide beamforming vector is achieved by tuning the position of the reference pinching antenna in the $k$-th waveguide. Specifically, based on \eqref{eq:waveguide_phase_align}, the position of the reference pinching antenna on the $k$-th waveguide can be expressed as 
\begin{equation}\label{eq:wave_Guide_reference_antenna_condition2}
\begin{split}
    {\phi_{k}} &= \left( {\frac{{2\pi }}{\lambda }\sqrt {\left\| {{{\bm{\varphi }}_{k,1}} - {\bm{u}}} \right\|}  + \frac{{2\pi }}{{{\lambda _G}}}\left\| {{{\bm{\varphi }}_{k,1}} - {{\bm{\varphi }}_{k,f}}} \right\|} \right)\\
    &= 2\tilde k\pi ,~~\tilde k \in Z.
\end{split}
\end{equation}

For clarity, we also provide an vanilla example for illustrating the relationship of \eqref{eq:wave_Guide_reference_antenna_condition2}, which is givn as follows.

\begin{example}
    Similar to the approach used to satisfy \eqref{eq:wave_Guide_reference_antenna_condition1}, the condition in \eqref{eq:wave_Guide_reference_antenna_condition2} is fulfilled by performing an antenna position search for each waveguide.  Revisiting \textbf{Example \ref{ex:waveguide_phase}},  we can now determine through this search that the reference antenna on Waveguide 1 is positioned at  ${\tilde \varphi _{1,1}} = \left( {19.975,0,10} \right)$, while that on  Waveguide 2  is located at ${\tilde \varphi _{2,1}} = \left( {19.935,30,10} \right)$.
\end{example}

Noting that ${\left\| {{{{\bm{\tilde h}}}_B}} \right\|^2}$ is the sum of squares of independent and identical distributed Gaussian random variables, which follows a Gamma Distribution $\Gamma(\nu,\theta)$ with shape parameter $\nu=N_B$ and scale parameter $\theta = \frac{1}{2 \sigma^2}=1$, we have $E\left\{ {{{{\bm{\tilde h}}}_B}} \right\} = {N_B}$. Following \eqref{eq:precoder_joint},  the average transmit power ratio of the BS over multiple fading channel cycles can be obtained as follows:
\begin{equation}\label{eq:PB_ratio_FCD}
{P_{F,B}} = \frac{{\frac{\eta }{{L_B^{\alpha }}}E\left\{ {{{\left\| {{{{\bm{\tilde h}}}_B}} \right\|}^2}} \right\}}}{{\frac{\eta }{{L_B^{\alpha }}}E\left\{ {{{\left\| {{{{\bm{\tilde h}}}_B}} \right\|}^2}} \right\} + \sum\limits_{k = 1}^K {\frac{{\eta {N_G}}}{{L_{G,k}^{\beta}}}} }} = \frac{{{N_B}}}{{{N_B} + {N_G}\sum\limits_{k = 1}^K {\frac{{L_B^{\alpha }}}{{L_{G,k}^{\beta}}}} }}.
\end{equation}
In this case, the $k$-th waveguide's average transmit power ratio can also be obtained as follows:
\begin{equation}\label{eq:PAS_ratio_FCD}
\begin{split}
{P_{F,G,k}} &= \frac{{\frac{{\eta {N_G}}}{{L_{G,k}^{\beta}}}}}{{\frac{{{N_B}\eta }}{{L_B^{\alpha }}}E\left\{ {{{\left\| {{{{\bm{\tilde h}}}_B}} \right\|}^2}} \right\} + \sum\limits_{k = 1}^K {\frac{{\eta {N_G}}}{{L_{G,k}^{\beta}}}} }} \hfill \\
&= \frac{{{N_G}}}{{{N_B}\frac{{L_{G,k}^{\beta}}}{{L_B^{\alpha }}} + {N_G}\sum\limits_{k' = 1}^K {\frac{{L_{G,k}^{\beta}}}{{L_{G,k'}^{\beta}}}} }}. \hfill \\
\end{split}
\end{equation}
We can observe from \eqref{eq:PB_ratio_FCD} and \eqref{eq:PAS_ratio_FCD} that the power allocation coefficients assigned to the BS increases as the number of BS antennas grows, i.e., $N_B$ is large, or when the BS is located closer to the UE relative to the PAS on the waveguides of interest, i.e., $L_B^{\alpha}$ is small. Otherwise, the waveguide will receive higher power allocation than the BS, if it has more PAS than   the antennas the BS have, i.e.,  when $N_G$ is large, or if the PAS  are closer to the UE than the BS, i.e., when $L_{G,k}^{\beta}$ is small.

Now the equivalent channel at the receiver can be  expressed as 
\begin{equation}
{h_F} = {\bm{h}}{{\bm{w}}_F} = \left\| {\bm{h}} \right\|.
\end{equation}
Therefore, the average channel gain in the full-cooperative scheme is given by 
\begin{equation}
\begin{gathered}
E\left\{ {{{\left| {{h_F}} \right|}^2}} \right\} = E\left\{ {\frac{\eta }{{L_B^{\alpha }}}{{\left\| {{{{\bm{\tilde h}}}_B}} \right\|}^2} + \sum\limits_{k = 1}^K {\frac{{\eta {N_G}}}{{L_{G,k}^{\beta}}}} } \right\} \hfill \\
= \frac{{\eta {N_B}}}{{L_B^{\alpha }}} + \sum\limits_{k = 1}^K {\frac{{\eta {N_G}}}{{L_{G,k}^{\beta}}}}.  \hfill \\ 
\end{gathered}
\end{equation}

According to \eqref{eq:SNR_express}, the average received SNR of the FCD scheme is 
\begin{equation}\label{eq:snr_FCD}
    \gamma_{FCD} = \frac{{\eta {N_B} P_t}}{{L_B^{\alpha } \sigma_n^2}} + \sum\limits_{k = 1}^K {\frac{{\eta {N_G} P_t}}{{L_{G,k}^{\beta} \sigma_n^2}}} .
\end{equation}

\subsection{Analysis and Discussion}

For ease of analysis and comparison, we assume uniform path loss across all waveguide antennas, i.e., $L_{G,k}^{\beta } = L_G^{\beta }$.  Let $\tilde \gamma = \frac{P_t}{\sigma_n^2}$ denote the transmit SNR. Table \ref{table:gathered_SNR} summaries the obtained analytical expression of the average received SNR for different BS-PAS joint transmission schemes.  We can see from Table \ref{table:gathered_SNR} that the received SNR of BS-PAS joint transmission depends on multiple factors, and the associated performance gains are not guaranteed for all parameter configurations. To this end, we analyze the achievable system gains with respect to different parameter settings.

\begin{table}
	\centering
	\caption{Average received SNR of different schemes}\label{table:gathered_SNR}
	\begin{tabular}{|c|c|}
		\hline
		Schemes & Average Received SNR Expression \\
		\hline
		BS-only & ${\gamma _{BO}} = \frac{{\eta {N_B}}}{{L_B^{\alpha }}}\tilde \gamma $  \\
		\hline
		BS+PAS SD & ${\gamma _{SD}} = \left( {\frac{{\eta N_B^2}}{{L_B^{\alpha }\left( {{N_B} + K} \right)}} + \frac{{\eta {N_G}K}}{{L_G^{\beta }\left( {{N_B} + K} \right)}}} \right)\tilde \gamma $ \\
		\hline
		BS+PAS SCD & ${\gamma _{SCD}} = \left( {\frac{{\eta N_B^2}}{{L_B^{\alpha }\left( {{N_B} + K} \right)}} + \frac{{\eta {N_G}{K^2}}}{{L_G^{\beta }\left( {{N_B} + K} \right)}}} \right)\tilde \gamma$ \\
		\hline
		BS+PAS FCD & ${\gamma _{FCD}} = \left( {\frac{{\eta {N_B}}}{{L_B^{\alpha }}} + \frac{{\eta {N_G}K}}{{L_G^{\beta }}}} \right)\tilde \gamma $ \\
		\hline
	\end{tabular}
\end{table}
To facilitate insightful analysis, we introduce the joint transmission gain to quantify the performance improvements of the BS-PAS joint transmission schemes relative to the BS-only scheme. Specifically, it is defined as the ratio of the average received SNR achieved by each of the three BS-PAS joint transmission schemes--SD,SCD, and FCD--to that of  the BS-only scheme.
Correspondingly, these gains are referred to as the  SD gain, SCD gain, and FCD gain, respectively, as shown in Table \ref{table:joint_transmission_Gain}. The analytical expressions for the proposed three schemes are provided in Table \ref{table:joint_transmission_Gain}.  In the following analysis, we  quantitatively examine  the impact of system parameters on  the gains of the different transmission schemes. By systematically varying relevant parameters in the gain expressions and analyzing the resulting trends, we can derive the following key insights.
\begin{table} 
	\centering
	\caption{Joint Transmission Gain of SD, SCD and FCD schemes}\label{table:joint_transmission_Gain}
	\begin{tabular}{|c|c|}
		\hline
		Joint Transmission Gain Ratio & Expressions\\
		\hline
		SD Gain ${V_{SD}} = \frac{{{\gamma _{SD}}}}{{{\gamma _{BO}}}}$ & $\frac{{{N_B}}}{{{{N_B} + K} }} + \frac{{{N_G}K}}{{{N_B}\left( {{N_B} + K} \right)}}\frac{{L_B^{\alpha }}}{{L_G^{\beta }}}$ \\
		\hline
		SCD Gain ${V_{SCD}} = \frac{{{\gamma _{SCD}}}}{{{\gamma _{BO}}}}$ & $\frac{{{N_B}}}{{ {{N_B} + K} }} + \frac{{{N_G}{K^2}}}{{{N_B}\left( {{N_B} + K} \right)}}\frac{{L_B^{\alpha }}}{{L_G^{\beta }}}$ \\
		\hline
		FCD Gain ${V_{FCD}} = \frac{{{\gamma _{FCD}}}}{{{\gamma _{BO}}}}$ & $1 + \frac{{{N_G}K}}{{{N_B}}}\frac{{L_B^{\alpha }}}{{L_G^{\beta }}}$ \\
		\hline
	\end{tabular}
\end{table}

\subsubsection{ Gains of the proposed joint transmission schemes} \label{para:gains_JDS}

As shown  in Table \ref{table:joint_transmission_Gain}, it is evident that the FCD scheme always provides a positive gain  of $\frac{{{N_G}K}}{{{N_B}}}\frac{{L_B^{\alpha }}}{{L_G^{\beta }}}$ compared to the BS-only scheme, since the FCD Gain ${V_{FCD}} > 1$ and $\frac{{{N_G}K}}{{{N_B}}}\frac{{L_B^{\alpha }}}{{L_G^{\beta }}}>0$ in  BS-PAS joint transmission systems. However, the SD and SCD schemes do not necessarily outperform the BS-only scheme. Their performance advantage depends on specific system parameters, such as the BS-UE path loss exponent $\alpha$, the number of BS antennas $N_B$, the number of waveguides $K$ and PAS $N_G$ and among others. 

\begin{remark}\label{re:gain_condition}
   Given  $V_{SD}>1$ and $V_{SCD}>1$, we have that 
   \begin{itemize}
       \item the condition for the SD scheme to outperform the BS-only scheme is $\frac{{L_B^{\alpha }}}{{L_G^{\beta }}} > \frac{{{N_B}}}{{{N_G}}}$, whereas 
       \item the SCD scheme achieves gain  over the BS-only scheme if $\frac{{L_B^{\alpha }}}{{L_G^{\beta }}} > \frac{{{N_B}}}{{{N_G}K}}$.
   \end{itemize}
\end{remark}
That is both the SD and SCD schemes outperform the BS-only scheme when $\frac{{L_B^{\alpha }}}{{L_G^{\beta }}} > \frac{{{N_B}}}{{{N_G}}}$. To illustrate the performance gain of SD and SCD schemes compared to BS-only in different scenarios, we provide an example as follows.
\begin{example}\label{ex:SD_SCD_Gain_condition}
    We use the typical scenario parameter values given in Table \ref{table:simulation_parameter} with varying $\alpha$. From the inequalities $\frac{{L_B^{\alpha }}}{{L_G^{\beta }}} > \frac{{{N_B}}}{{{N_G}}}$ and $\frac{{L_B^{\alpha }}}{{L_G^{\beta }}} > \frac{{{N_B}}}{{{N_G}K}}$, we can compute that ${\alpha} > 2.13$ and ${\alpha} > 1.87$, respectively. This indicates that in the considered scenario, the SD scheme outperforms the BS-only scheme when $\alpha > 2.13$, while the SCD scheme always outperforms BS-only scheme as $\alpha  \geqslant 2 $. 
\end{example}

\begin{example}\label{ex:SD_SCD_Gain_condition_N_G}
    From \textbf{Remark \ref{re:gain_condition}}, the conditions under which SD and SCD schemes outperform the BS-only scheme are given by ${N_G} > \frac{{{N_B}L_G^{\beta }}}{{L_B^{\alpha }}}$  and ${N_G} > \frac{{{N_B}L_G^{\beta }}}{{KL_B^{\alpha }}}$, respectively.  Using the typical  parameter values given in Table \ref{table:simulation_parameter} with $\alpha = \beta = 2$, we obtain the thresholds  ${N_G} > 16$ for the SD scheme and ${N_G} > 4$ for the SCD scheme, respectively.
\end{example}

\subsubsection{Impact of path loss} 
We can observe from Table \ref{table:joint_transmission_Gain} that the gains of SD, SCD and FCD schemes over the BS-only scheme monotonically relies on the ratio of the path loss  between the BS-UE and the PAS-UE. That is, the gains attained by the proposed transmission schemes grows when the ratio of the path loss, i.e.,  $\frac{{L_B^{\alpha }}}{{L_G^{\beta }}}$, increases. This indicates that cooperation with PAS can provide additional large-scale channel gains. 

\begin{remark}\label{re:BS-UEexponent}
    ${V_{FCD}} = 1$ when ${L_B^{\alpha }} \ll {L_G^{\beta }}$, i.e., $\frac{{L_B^{\alpha }}}{{L_G^{\beta }}}$ goes to $0$. 
\end{remark}
This means that under favorable  BS-UE propagation conditions--characterized by a low path loss exponent or short transmission distance--the FCD and BS-only schemes yield comparable performance. In such scenario, the dominant contribution comes from the strong direct BS-UE link.

\begin{remark} \label{re:extra_cooperative_gain}
It should be noticed from Table \ref{table:joint_transmission_Gain} that  $\frac{{{V_{SCD}}}}{{{V_{SD}}}} = K$ and  $\frac{{{V_{FCD}}}}{{{V_{SCD}}}} = 1 + \frac{{{N_B}}}{K}$, when $L_B^{\alpha } \gg L_G^{\beta }$, i.e., $\frac{{L_B^{\alpha }}}{{L_G^{\beta }}}$ approaches infinity.
\begin{proof}
When ${\frac{{L_B^{\alpha }}}{{L_G^{\beta }}}} \to \infty $, the first term ${\frac{{{N_B}}}{{{{N_B} + K} }}}$ in the expression of  ${{V_{SD}}}$ and ${{V_{SCD}}}$ can be omitted. By dividing $V_{SCD}$ by ${V_{SD}}$, we have
\begin{equation}
\mathop {\lim }\limits_{\frac{{L_B^{\alpha }}}{{L_G^{\beta }}} \to \infty } \frac{{{V_{SCD}}}}{{{V_{SD}}}} = \frac{{\frac{{{N_G}{K^2}}}{{{N_B}\left( {{N_B} + K} \right)}}\frac{{L_B^{\alpha }}}{{L_G^{\beta }}}}}{{\frac{{{N_G}K}}{{{N_B}\left( {{N_B} + K} \right)}}\frac{{L_B^{\alpha }}}{{L_G^{\beta }}}}} = K.
\end{equation}

Furthermore, the first term ${\frac{{{N_B}}}{{\left( {{N_B} + K} \right)}}}$ in $V_{SCD}$ and the constant term $1$ in $V_{FCD}$ can be omitted if ${\frac{{L_B^{\alpha }}}{{L_G^{\beta }}}} \to \infty $. Similarly, by dividing $V_{FCD}$ by ${V_{SCD}}$, we have
\begin{equation}
\begin{gathered}
\mathop {\lim }\limits_{\frac{{L_B^{\alpha }}}{{L_G^{\beta }}} \to \infty } \frac{{{V_{FCD}}}}{{{V_{SCD}}}} = \frac{{\frac{{{N_G}K}}{{{N_B}}}\frac{{L_B^{\alpha }}}{{L_G^{\beta }}}}}{{\frac{{{N_G}{K^2}}}{{{N_B}\left( {{N_B} + K} \right)}}\frac{{L_B^{\alpha }}}{{L_G^{\beta }}}}} = 1 + \frac{{{N_B}}}{K}. \hfill \\
\end{gathered} 
\end{equation}
\end{proof}
\end{remark}
 The SD scheme processes each waveguide independently, resulting in comparatively lower gains due to the absence of inter-waveguide collaboration. \textbf{Remark \ref{re:extra_cooperative_gain}} shows that the $K$-fold gain of the SCD scheme over the SD scheme, which is achieved through coordination among the $K$ waveguides. Moreover, the performance gain of  the FCD  scheme over the  SCD scheme comes from the additional coordination between the BS and the $K$ waveguides. Since full cooperation in the FCD scheme can avoid allocating power to the BS antennas with poor channel conditions, its the performance gain over the SCD scheme increases with the number of BS antennas.

\subsubsection{Impact of the number of BS antennas}
From Table \ref{table:gathered_SNR}, it can be observed that since $N_B$ only appears in the numerator of the received SNR for both the BS-only and FCD schemes, their performance improves monotonically as the number of base station antennas increases. However, for the SD and SCD schemes, the two terms of received SNR vary in opposite directions with increasing $N_B$, resulting in potentially non-monotonic performance behaviors.

\begin{remark}\label{re:Infinite_BS_antennas}
    As the number of base station antennas approaches infinity, it can be observed from Table \ref{table:joint_transmission_Gain} that ${V_{SD}}$,$V_{SCD}$ and $V_{FCD}$ all asymptotically converge to $1$. That is, all the three joint transmission schemes achieve the same performance as the BS-only scheme. 
\end{remark}
In this asymptotic case, nearly all transmit power is allocated to the BS, thereby rendering PAS ineffective. However, the performance of the three joint transmission schemes converge to that of the BS-only scheme at different rates. 

\begin{example} \label{ex:Infinite_BS_antennas_V_FCD_3dB}
    Using the set of typical scenario parameters in Table \ref{table:simulation_parameter} and the gain expression of FCD scheme in Table \ref{table:joint_transmission_Gain},  we can compute that  ${V_{FCD}} \approx 1 + \frac{{1224}}{{{N_B}}}$. 
\end{example}
This result indicates that when the number of BS antennas reaches approximately to $1224$, the gain of the FCD over the BS-only scheme reduces to 3dB. Therefore, in typical scenarios, the FCD scheme consistently offers substantial performance improvements.

\subsubsection{Impact of the number of waveguides}

Now we turn to investigate the impact of the number of waveguides by analyzing the expressions in Table \ref{table:joint_transmission_Gain}. First, as the number of waveguides increases, the performance gain of the joint transmission schemes over the BS-only scheme increase. 
\begin{remark}\label{re:infinite_waveguides}
    As the number of waveguides $K$ approaches infinity, we find that the joint transmission gain becomes $V_{SD} = \frac{{N_G}{L_B^{\alpha }}}{{{N_B}L_G^{\beta }}}$ and $V_{SCD} = V_{FCD} = \frac{K{N_G}{L_B^{\alpha }}}{{{N_B}L_G^{\beta }}}$.
\end{remark}
On the one hand, the gain of the SD scheme converges to a constant value, since nearly all power is allocated to PAS, but there is no array gain due to the lack of cooperation among different waveguides. On the other hand, both SCD and FCD schemes achieve gains that increase linearly with the number of waveguides $K$.

Furthermore, it should be pointed out that as $K$ increases, the received SNR at the UE becomes increasingly dependent on the transmission of PAS. Consequently, the performance of the SCD scheme asymptotically approaches that of the FCD scheme. The performance gain of the FCD scheme over the SCD scheme can be given by
\begin{equation}\label{eq:VFCDvsVSCD}
\frac{{{V_{FCD}}}}{{{V_{SCD}}}} = 1 + \frac{{{N_B}K + {N_B}K{N_G}\frac{{L_B^{\alpha }}}{{L_G^{\beta }}}}}{{N_B^2 + {K^2}{N_G}\frac{{L_B^{\alpha }}}{{L_G^{\beta }}}}}.
\end{equation}
Suppose $K=N_B$, then the gain of the SCD scheme over the SCD scheme is $\frac{{{V_{FCD}}}}{{{V_{SCD}}}} = 2$, i.e., a 3dB improvement. Since the PAS typically are activated on limited waveguides, the FCD scheme generally outperforms the SCD scheme in most practical scenarios.

\subsubsection{Impact of the number of PAS}

Finally, we investigate how the number of PAS affects the performance gains  of the three proposed joint transmission schemes compared to the BS-only scheme. It can be observed from Table \ref{table:joint_transmission_Gain} that the number of PAS on each waveguide $N_G$ only appears at the numerator of the expression of $V_{SD},V_{SCD}$  and $V_{FCD}$, thereby  the performance gain of joint transmission schemes over the BS-only scheme increase linearly with $N_G$. 

However, if the number of PAS is small, the expression of received SNR of the SD and SCD scheme may be smaller than $1$, that means too few PAS on PAS make the SD and SCD schemes inferior to the BS-only scheme. This is because the waveguide diverts part of the transmit power from BS, but the insufficient number of PAS results in inadequate beamforming gain.

\section{Simulation Results}
In this section, we perform Monte Carlo simulations and numerical evaluations for estimating the performance of the three proposed  BS-PAS joint transmission schemes, i.e., SD, SCD, and FCD.  For comparison, we consider the conventional BS-only scenario, which serves as the benchmark. In the following simulations, we firstly verify the correctness of our theoretical derivations. Then, we evaluating the achievable gains of the proposed three BS-PAS joint transmission schemes with respect to various system parameters, to offer practical insights practical insights for the design and deployment of future wireless communication networks.  
Unless otherwise specified,  the values of parameters provided in Table \ref{table:simulation_parameter} are used throughout the following simulations. In this paper, we consider the typical scenario where the BS-UE path loss is  greater than that of the PAS, i.e., the path loss exponents in Table \ref{table:simulation_parameter} satisfy  $\alpha  \geqslant  \beta$. This assumption is due to the fact that the path-loss exponent of an NLoS link is greater that of an LoS link.

\begin{table} 
	\centering
	\caption{Simulation Parameters}\label{table:simulation_parameter}
	\begin{tabular}{|c|c|}
		\hline
		Parameters & Value \\
		\hline
		Carrier Frequency & $3.5$ GHz  \\
		\hline
		Bandwidth & $100$ MHz \\
		\hline
		Noise Power Density & $-170$ dBm/Hz \\
		\hline
		Transmit Power & $1  \sim  100$ W ($0 \sim 20$ dB)  \\
		\hline
		BS-UE Distance & $L_B=200$m \\
		\hline
		RPA-User Distance  & $L_G = 100$ m \\
		\hline
		Number of BS Antennas & $N_B = 64$ \\
		\hline
		Number of PAS & $N_G = 8$ \\
		\hline
		Number of Waveguides  & $K = 4$ \\
		\hline
		Path loss exponents & $\alpha=2.4$, $\beta = 2$\\
		\hline
		Monte Carlo Simulation Number & $10000$ \\
		\hline
	\end{tabular}
\end{table}

We start by simulating the average SNR received at the UE achieved by the SD, SCD and BS-only schemes. Specifically, in the considered scenario, the BS-UE channel follows a  Rayleigh fading model with i.i.d complex Gaussian coefficients $\mathcal{CN}(0,1)$ with unit variance.  The distance between the UE and the BS is set to $200$ meters. Suppose that the actual distance between the UE and the reference pinching antenna of a waveguide is randomly generated within the range $\left[ 100- \frac{\lambda}{2}, 100 + \frac{\lambda}{2} \right] $, ensuring that the phase delay of signals arriving at the UE via PAS on each waveguide follows a uniform distribution over $[0,2\pi]$.

\begin{figure}
	\vspace{-0.1cm}
	\centering
	\includegraphics[width=0.4\textwidth]{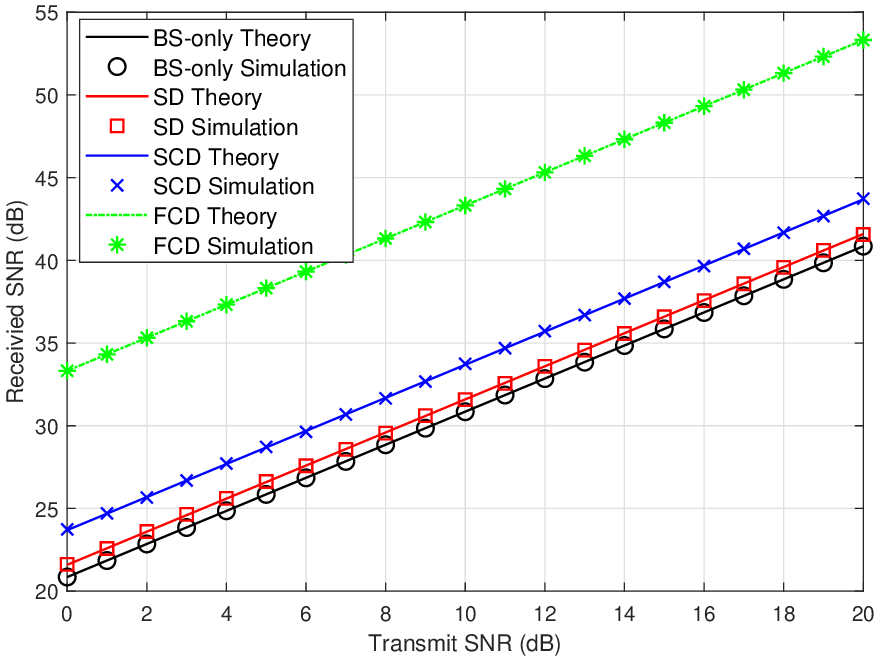}
	\caption{Average received SNR of different schemes.}\label{fig:SNR_validate}
    \vspace{-1.5em}
\end{figure}

Fig. \ref{fig:SNR_validate} shows the average SNR received at the UE versus transmit power for the four different schemes considered.  We  see that the simulated SNR matches well with the analytical curves derived in  closed-form expressions in \eqref{eq:snr_SD},~\eqref{eq:snr_SCD} and \eqref{eq:snr_FCD}. We can also observe from Fig. \ref{fig:SNR_validate} that the three proposed BS-PAS joint transmission schemes gradually improve the average SNR received at the UE compared to the BS-only scheme, with the FCD scheme achieving the best performance among them. These results follow from  the expressions presented in Table \ref{table:gathered_SNR}, and further details on how various factors influence the performance gain will be illustrated in the subsequent simulation results.

\begin{figure}
	\vspace{-0.1cm}
	\centering
	\includegraphics[width=0.4\textwidth]{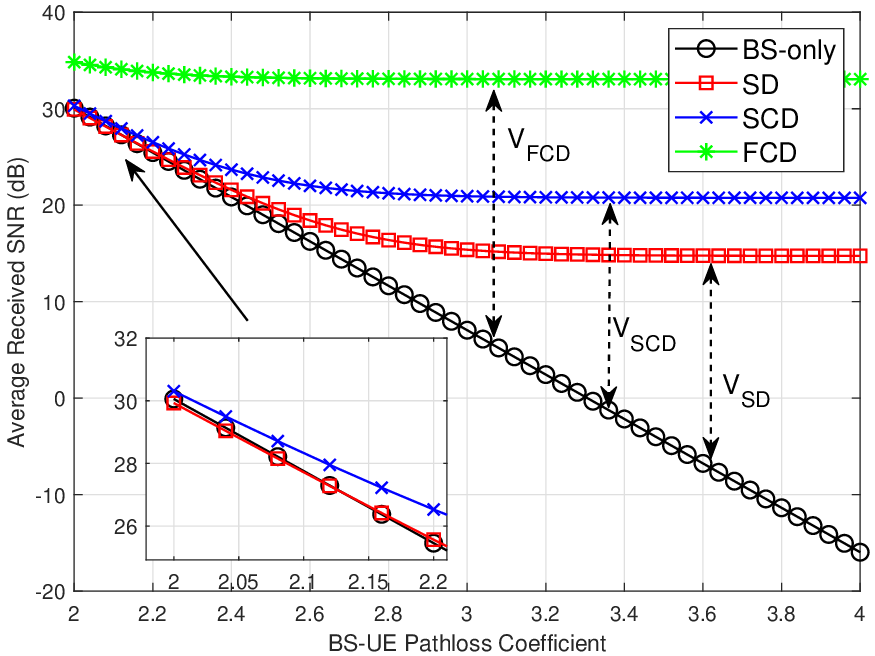}
	\caption{Average received SNR versus different BS-UE path loss exponents.}\label{fig:AverageSNRPathlossCoef}
    \vspace{-1.5em}
\end{figure}

Fig. \ref{fig:AverageSNRPathlossCoef} shows the impact of  the BS propagation environment on the performance of the three proposed joint transmission schemes and their respective gains compared to the BS-only scheme.  Here, $\beta = 2$ and the value of $\alpha$ is varied from $2$ to $4$. This allows us to evaluate how changes in the BS propagation environment influence the performance of the proposed three joint transmission schemes and their gains relative to the BS-only scheme.

 We can see from the inset of Fig. \ref{fig:AverageSNRPathlossCoef} that the SD scheme is inferior to the BS-only scheme when the BS-UE path loss exponent falls below $2.13$, while the SCD scheme always outperforms the BS-only scheme. This result is consistent with the gain conditions for SD, SCD outlined in \textbf{Remark \ref{re:gain_condition}} and \textbf{Example \ref{ex:SD_SCD_Gain_condition}}. Moreover, as the BS-UE path loss exponent continues to decrease, the performance of the FCD and BS-only schemes converges to the same value. This observation is consistent with findings discussed in \textbf{Remark \ref{re:BS-UEexponent}}.

We can see from Fig. \ref{fig:AverageSNRPathlossCoef} that the average received SNR of all schemes decreases with increasing the path loss exponent between the BS and the UE. Meanwhile, the FCD scheme consistently attains the best average received SNR compared to the other schemes across the entire range of BS-UE path loss exponents. Moreover, Fig. \ref{fig:AverageSNRPathlossCoef} also shows that the performance gap between the three proposed joint transmission schemes and the BS-only scheme widens as the BS-UE path loss exponent increases. Specifically, the received SNR of the BS-only scheme continuously decreases with increasing the BS path loss exponent, while the three joint transmission schemes converge to fixed performance levels. 
Furthermore, as the BS-UE path loss coefficient  $\alpha$ increases, the additional performance gain of SCD over SD arising from cooperation between different waveguides, and the additional performance gain of FCD over SCD enabled by cooperation between BS and waveguides, will converge to fixed values as described in \textbf{Remark \ref{re:extra_cooperative_gain}}.

\begin{figure}
	\vspace{-0.1cm}
	\centering
	\includegraphics[width=0.4\textwidth]{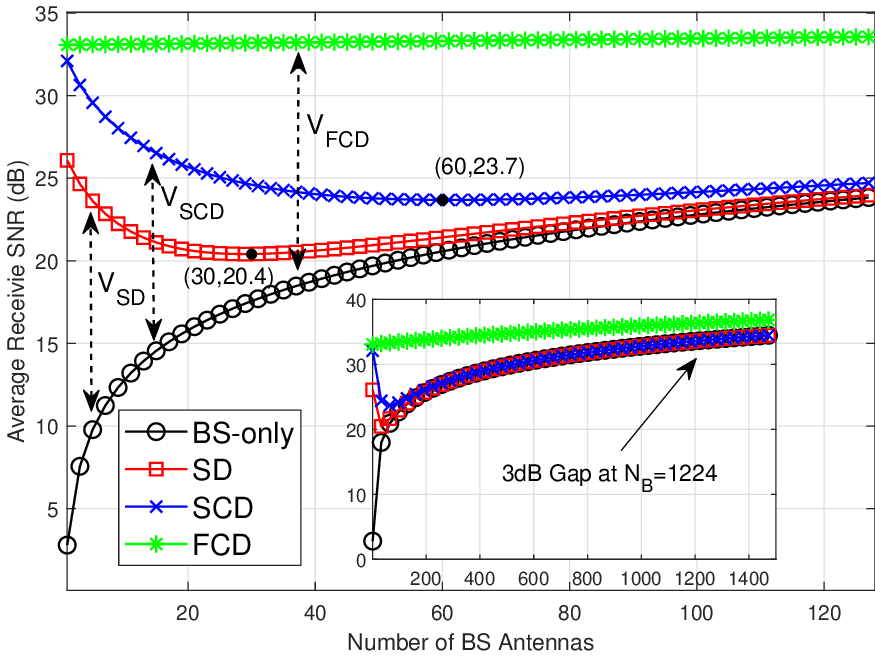}
	\caption{Average received SNR with different number of BS antennas $N_B$.}\label{fig:AverageSNRBsAntennas}
    \vspace{-1.5em}
\end{figure}

Fig. \ref{fig:AverageSNRBsAntennas} illustrates the impact of the number of BS antennas on  the received SNR across different schemes, where the number of BS antennas ranges from $1$ to $128$. 
We see that the performance of both the BS-only and FCD schemes monotonically increases as  the number of BS antennas grows. Nevertheless, as $\alpha > \beta$, the received signal power is primarily dominated by contributions from the PAS. Consequently, the performance of the FCD scheme exhibits  a slower rate of improvements  as  $N_B$ increases compared to the other schemes.

For the SD and SCD schemes in Fig. \ref{fig:AverageSNRBsAntennas}, the received SNR initially decreases and then increases as the number of BS antennas grows. 
When the BS has a small number of antennas, increasing the number of BS antennas diverts more power away from the PAS, resulting in performance degradation. However, when the number of BS antennas reaches a certain threshold, e.g., $30$ for SD and $80$ for SCD, the beamforming gain becomes sufficient to compensate for the BS's path loss disadvantage, allowing the SD and SCD schemes to outperform BS-only scheme. Finally, the inset of Fig. \ref{fig:AverageSNRBsAntennas} shows that the gap between the BS-only, SD and SCD schemes narrows as the number of BS antennas increases, which aligns with the findings in \textbf{Remark \ref{re:Infinite_BS_antennas}}. Moreover, it can be calculated that the performance gap between the FCD scheme and the BS-only scheme narrows to about $3$dB when the number of BS antennas reaches $1224$, as shown in \textbf{Example \ref{ex:Infinite_BS_antennas_V_FCD_3dB}}.

\begin{figure}[t]
	\centering
	\subfigure[$\alpha=2.4,\beta=2$.]{
		\label{fig:AverageSNRWaveguideNumLargeBSPathloss}
		\includegraphics[width=0.4\textwidth]{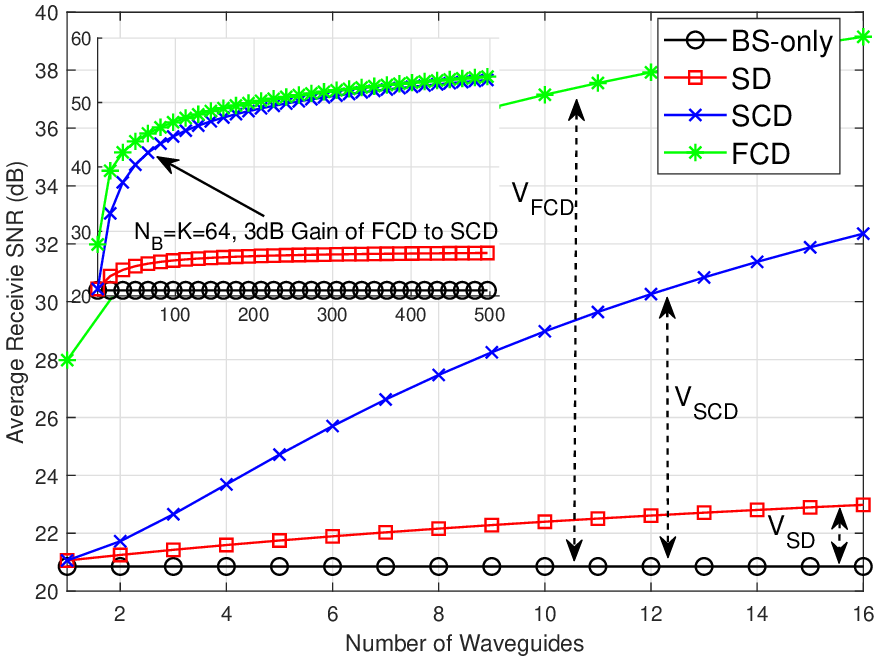}
	} \\
	\subfigure[$\alpha=2,\beta=2$.]{
		\label{fig:AverageSNRWaveguideNumSmallBSPathloss}
		\includegraphics[width=0.4\textwidth]{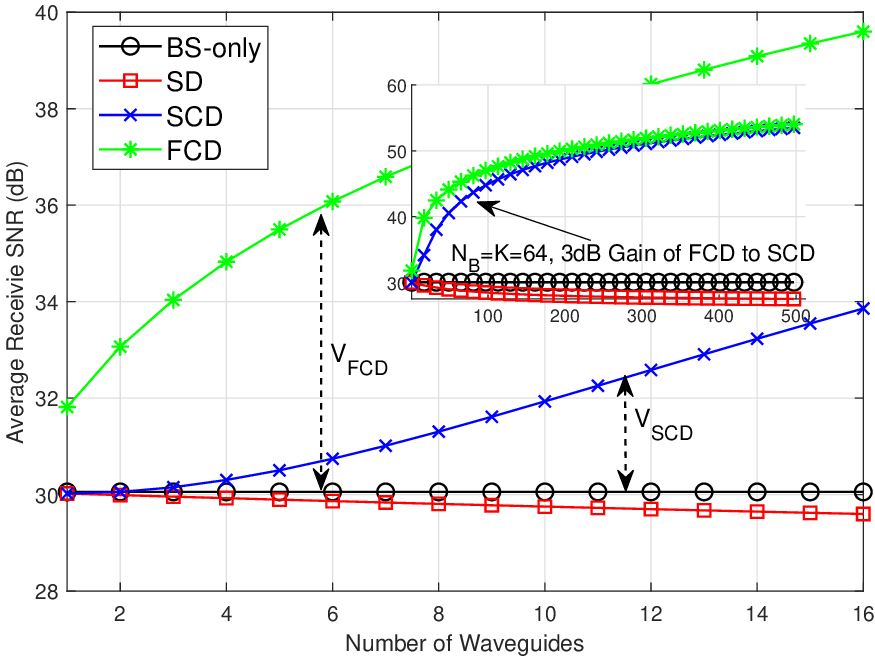}
	}
	\caption{Average received SNR versus different numbers of waveguides for different transmission schemes.}\label{fig:AverageSNRWaveguideNum}
    \vspace{-1.5em}
\end{figure}

Fig. \ref{fig:AverageSNRWaveguideNum} demonstrates how the number of waveguides affects the system performance. To illustrate the benefits of employing PAS in different communication scenarios, two different BS-UE path loss exponents, $\alpha=2.4$ and $\alpha=2$, are used in Fig. \ref{fig:AverageSNRWaveguideNumLargeBSPathloss} and Fig.  \ref{fig:AverageSNRWaveguideNumSmallBSPathloss}, respectively. 
As shown in Fig. \ref{fig:AverageSNRWaveguideNumLargeBSPathloss}, all the three proposed joint transmission schemes outperform the BS-only scheme. As the number of waveguides increases, the performance gain of the proposed joint transmission schemes over the BS-only scheme exhibits a rapid initial growth. 
As the number of waveguides increases further, the gain trends of the three joint transmission schemes begin to diverge. Specifically, the gain of the SD scheme converges to a constant value, while the performance gains of both the SCD and FCD schemes increase monotonically with the number of waveguides $K$, eventually  asymptotically converging to the same level. These results are consistent with the findings in \textbf{Remark \ref{re:infinite_waveguides}.}
In Fig. \ref{fig:AverageSNRWaveguideNumSmallBSPathloss},  the BS and PAS share the same path loss exponent. the SD scheme demonstrates no gain over BS-only, because $\alpha=2<2.13$ and the condition in \textbf{Example \ref{ex:SD_SCD_Gain_condition}} is not satisfied. 
Furthermore, the insets in Fig. \ref{fig:AverageSNRWaveguideNumSmallBSPathloss} and Fig. \ref{fig:AverageSNRWaveguideNumSmallBSPathloss} illustrates the gain for the FCD scheme over the SCD scheme decreases to 3dB when $K$ increase to $64$, i.e., equals to $N_B$, which are consistent to the analysis under \textbf{Remark \ref{re:infinite_waveguides}}.

\begin{figure}[t]
	\centering
	\subfigure[$\alpha=2.4,\beta=2$.]{
		\label{fig:AverageSNRPinchingAntennasLargeBsPathloss}
		\includegraphics[width=0.4\textwidth]{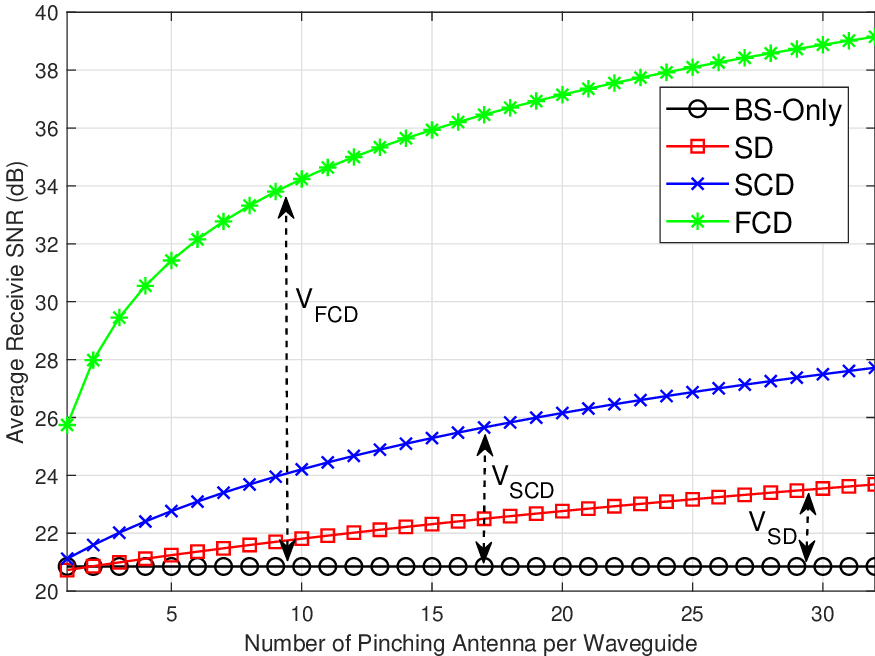}
	} \\
	\subfigure[$\alpha=2,\beta=2$.]{
		\label{fig:AverageSNRPinchingAntennasSmallBsPathloss}
		\includegraphics[width=0.4\textwidth]{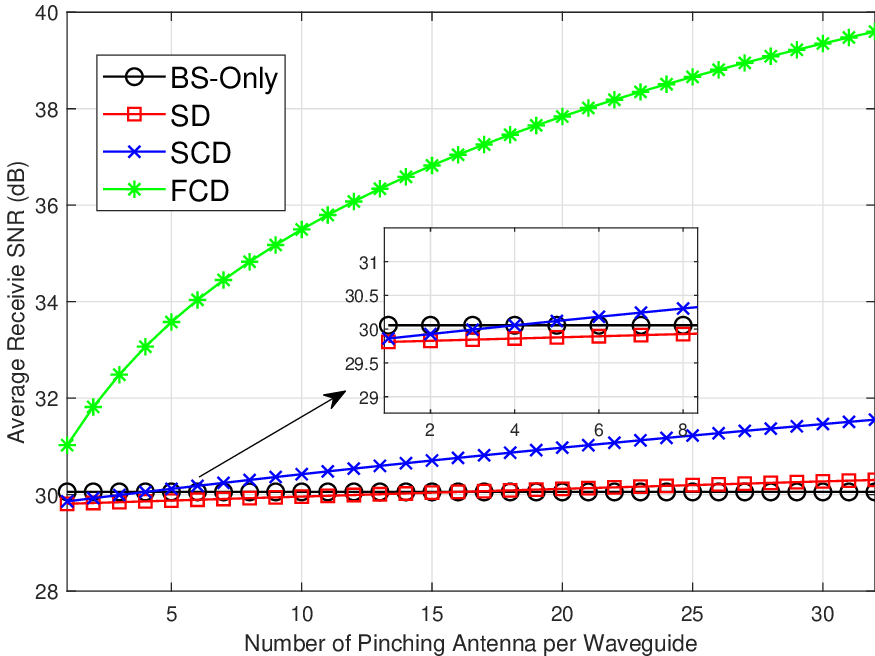}
	}
	\caption{Average received SNR with different number of PAS on a waveguide.}\label{fig:AverageSNRPinchingAntennaNum}
    \vspace{-1.5em}
\end{figure}

Fig. \ref{fig:AverageSNRPinchingAntennaNum} illustrates the average received SNR performance versus the number of PAS distributed on each waveguide, with two different BS-UE path loss exponents, i.e., $\alpha = 2.4$ and $\alpha=2$. 
It shows that the performance gain of joint transmission schemes over the BS-only scheme increase consistently with the number of PAS on each waveguide. 
Moreover, we can see from the inset in Fig. \ref{fig:AverageSNRPinchingAntennasSmallBsPathloss} that the SD and SCD schemes initially exhibit poorer performance than the BS-only scheme, since the conditions in \textbf{Example \ref{ex:SD_SCD_Gain_condition_N_G}} are not satisfied. When the number of pinch antennas on each waveguide increases to $16$ and $4$ for SD and SCD, respectively, that is, conditions in \textbf{Example \ref{ex:SD_SCD_Gain_condition_N_G}} are satisfied, the SD and SCD schemes eventually achieve performance gains. 
 
\section{Conclusions and Future Works}
This paper has investigated joint transmission architectures between BS and PAS, aiming for addressing the deployment challenges of PAS in cellular networks.  Specifically,  we proposed three different joint transmission schemes, namely SD, SCD and FCD based on the cooperation level between  the BS and the PAS. We also developed  optimal beamforming schemes based on the MRT principle  and  derived closed-form expressions of the average received SNR for each transmission strategy. Simulation results were provided to reveal that  proposed cooperative transmission strategies between the BS and PAS  can significantly enhance communication performance. Moreover, by analyzing the impact of key system parameters, e.g., path loss exponents and the number of PAS, on the transmission efficiency, this work sheds light on the practical deployment of cellular systems integrated with PAS.

Nevertheless, the deployment of PAS in cellular networks presents several key challenges and highlights important directions for future research. 
For example, efficient coordination between the BS and PAS must be established to support multi-user spatial multiplexing. Additionally, further efforts are needed to reduce the overhead of deploying PAS while minimizing modifications to the existing BS infrastructure. Another key challenge involves joint time-frequency resource scheduling between the BS and PAS. These open issues present promising directions for future research.

\vspace{-0.1cm}

\appendices
\section{Proof of \textbf{Proposition \ref{prop:ave-channel-gain-SD}}}
\label{proof-prop:ave-channel-gain-SD}
In this section, we derive the average channel gain for the SD scheme in detail. Since the distance between the reference pinching antenna of each waveguide and the UE is random, we assume that  $\phi_k$ follows a uniform distribution of $[0,2\pi]$.
For clarity, let 
\begin{equation}\label{eq:waveguides_channel_vector_sum}
G{e^{ - j\Omega }} = \sum\limits_{k = 1}^K \sqrt{{\frac{1}{{{L_{G,k}^\beta}}}}{e^{ - j{\phi _k}}}}.
\end{equation}

Now, we have that the channel gain can be expressed as follows:
\begin{equation}\label{eq:standaloe_channel_power_Gain_derive}
\begin{gathered}
E\left\{ {{{\left| {{h_S}} \right|}^2}} \right\} = E\left\{ {{{\left| {\sqrt{\frac{\eta {P_{S,B}}}{{L_B^\alpha }}} \left\| {{{{\bm{\tilde h}}}_B}} \right\| + \sqrt {\eta {N_G}{P_{S,G}}} G{e^{ - j\Omega }}} \right|}^2}} \right\} \hfill \\
= E\left\{ {\frac{{\eta {P_{S,B}}}}{{L_B^{\alpha }}}{{\left\| {{{{\bm{\tilde h}}}_B}} \right\|}^2}} \right\} \hfill \\
 + E\left\{\eta \sqrt{\frac{{{P_{S,B}}{N_G}{P_{S,G}}} }{{L_B^\alpha }}} \left\| {{{{\bm{\tilde h}}}_B}} \right\|G\cos \left( \Omega  \right) \right\} \hfill\\
+ E\left\{ {\eta {N_G}{P_{S,B}}{G^2}} \right\}. \hfill  \\
\end{gathered} 
\end{equation}

For the first term of \eqref{eq:standaloe_channel_power_Gain_derive}, it is known that $E\left\{ {{{\left\| {{{{\bm{\tilde h}}}_B}} \right\|}^2}} \right\} = {N_B}$ with ${{{{\bm{\tilde h}}}_B}}$ being the normalized channel coefficient, we have
\begin{equation}\label{eq:standalone_first_term}
E\left\{ {\frac{{\eta {P_{S,B}}}}{{L_B^{\alpha }}}{{\left\| {{{{\bm{\tilde h}}}_B}} \right\|}^2}} \right\} = \frac{{\eta {N_B}{P_{S,B}}}}{{L_B^{\alpha }}}.
\end{equation}

For the second term  of \eqref{eq:standaloe_channel_power_Gain_derive}, we can prove that $\Omega$ follows a uniform distribution over $\left(0,2\pi \right) $, i.e., $\Omega \sim U\left( 0, 2\pi \right) $, as follows. 

\begin{enumerate}

\item  Given $a_1$, $a_2$, and two random variables ${\omega _1}$ and $\omega_2$, which both follow uniform distributions over $\left( 0, 2\pi \right) $, we can define a random variable (RV) $ G_1 {e^{j \Omega_1 }} = {a_1}{e^{j{\omega _1}}} + {a_2}{e^{j{\omega _2}}}$. Then we can define the RV $\hat \Omega_1 = \bmod \left( {{{ \Omega }_1} ,2\pi } \right)$ with support on $\left( {0,2\pi } \right)$. \label{step-1}

\item Based on the properties of vector addition in a two-dimensional plane, we can conclude that the conditional probability density function (CPDF) of $\hat \Omega_1$ depends on the absolute difference between $\hat \Omega_1$ and $\omega_1$, thus its CPDF with $\omega_1$ can be denoted as ${f_{{{\hat \Omega }_1},{\omega _1}}}\left( \Phi  \right) = F \left( {\Delta \Phi } \right)$, where $\Delta \Phi  = \bmod \left( {\Phi  - {\omega _1},2\pi } \right)$.  \label{step-2}

\item Note that for any given $\Phi$, the range of $\Delta \Phi$ is $\left[ 0, 2\pi\right]$ when $\omega_1$ varies in $\left[ 0, 2\pi\right]$. Based on the definition of PDF, we have $\int\limits_0^{2\pi } {F\left( {\Delta \Phi } \right)} d\Phi  = \int\limits_0^{2\pi } {F\left( {\Delta \Phi } \right)} d{\omega _1} = 1$. Since $\omega_1 \sim U\left(0,2\pi\right)$, the PDF of $\hat \Omega_1$ can be given by 
\begin{equation}
\begin{gathered}
  {f_{{{\hat \Omega }_1}}}\left( \Phi  \right) = \int\limits_0^{2\pi } {{f_{{{\hat \Omega }_1},{\omega _1}}}\left( \Phi  \right)\frac{1}{{2\pi }}d{\omega _1}}  \hfill \\
 = \frac{1}{{2\pi }}\int\limits_0^{2\pi } {F \left( {\Delta \Phi } \right)d{\omega _1}} = \frac{1}{{2\pi }}.  \hfill \\ 
\end{gathered} 
\end{equation} 
We thus proved that $\bmod \left( {\Omega_1 ,2\pi } \right)$ follows a uniform distribution over $\left( {0,2\pi } \right)$. \label{step-3}

\item  Set ${ G_k}{e^{j{{ \Omega }_k}}} = { G_{k-1}}{e^{j{{ \Omega }_{k-1}}}} + {a_{k+1}}{e^{j{\omega_{k+1}}}}$ and repeat the above proof in step \ref{step-1} $\sim$ \ref{step-3} for $k=2,\ldots,K-1$. We can prove that ${G_{K - 1}}{e^{j{\Omega _{K - 1}}}} = \sum\nolimits_{k = 1}^K {{a_k}{e^{j{\omega _k}}}} $ follows a uniform distribution over $\left( {0,2\pi } \right)$.  \label{step-4} 
	
\item Let ${a_k} = \sqrt{\frac{1}{{{L_{G,k}^\beta}}}}$ and ${\omega _k} =  - {\phi _k}$. Combining with \eqref{eq:waveguides_channel_vector_sum}, it is proved that $\Omega$ follows a uniform distribution over $\left( {0,2\pi } \right)$.  \label{step-6}
\end{enumerate}

Therefore, we have that $E\left\{ {\cos \left( \Omega  \right)} \right\}=0$. Furthermore, note that  ${\left\| {{{{\bm{\tilde h}}}_B}} \right\|}$, $G$, and ${\cos \left( \Omega  \right)}$ are uncorrelated, thus the second term of \eqref{eq:standaloe_channel_power_Gain_derive} is zero.

Now we move to the third term of \eqref{eq:standaloe_channel_power_Gain_derive}.  We start by proving the following equation:
\begin{equation}
E\left\{ {{G^2}} \right\} = \sum\nolimits_{k = 1}^K {\frac{1}{{L_{G,k}^{\beta}}}}
\end{equation}
which can be  derived following the below calculations. 
{\small
\begin{align}
&G{e^{ - j\Omega }} = \sum\limits_{k = 1}^K {\sqrt{\frac{1}{{{L_{G,k}^\beta}}}}\cos \left( {{\phi _k}} \right)}  - i\sum\limits_{k = 1}^K {\sqrt{\frac{1}{{{L_{G,k}^\beta}}}}\sin \left( {{\phi _k}} \right)}  \hfill, \\
&{G^2} = {\left( {\sum\limits_{k = 1}^K {\sqrt\frac{1}{{{L_{G,k}^\beta}}}\cos \left( {{\phi _k}} \right)} } \right)^2} + {\left( {\sum\limits_{k = 1}^K {\sqrt\frac{1}{{{L_{G,k}^\beta}}}\sin \left( {{\phi _k}} \right)} } \right)^2} \hfill ,\\
&E\left\{ {{G^2}} \right\} = \sum\limits_{k = 1}^K {E\left\{ {\frac{1}{{L_{G,k}^{\beta}}}\cos {{\left( {{\phi _k}} \right)}^2}} \right\}}  + \sum\limits_{k = 1}^K {E\left\{ {\frac{1}{{L_{G,k}^{\beta}}}\sin {{\left( {{\phi _k}} \right)}^2}} \right\}}  \hfill ,\\
&E\left\{ {{G^2}} \right\} = \frac{1}{2}\sum\limits_{k = 1}^K {\frac{1}{{L_{G,k}^{\beta}}}}  + \frac{1}{2}\sum\limits_{k = 1}^K {\frac{1}{{L_{G,k}^{\beta}}} = } \sum\limits_{k = 1}^K {\frac{1}{{L_{G,k}^{\beta}}}}  \hfill ,\label{eq:G_square}
\end{align}
}
where \eqref{eq:G_square} is obtained follows from $\phi_k \sim U\left( 0, 2\pi\right) $. Therefore, the third term of \eqref{eq:standaloe_channel_power_Gain_derive} is given by
\begin{equation}\label{eq:standalone_third_term}
E\left\{ {\frac{{\eta {N_G}}}{{\left( {{N_B} + K} \right)}}{G^2}} \right\} = \frac{{\eta {N_G}}}{{\left( {{N_B} + K} \right)}}\sum\limits_{k = 1}^K {\frac{1}{{L_{G,k}^{\beta}}}} .
\end{equation}

By summing \eqref{eq:standalone_first_term} and \eqref{eq:standalone_third_term}, we can obtain
\begin{equation}\label{eq:standalone_channel_power_Gain_temp}
E\left\{ {{{\left| {{h_S}} \right|}^2}} \right\} = \frac{{\eta {N_B}{P_{S,B}}}}{{L_B^{\alpha }}} + \eta {N_G}{P_{S,G}}\sum\limits_{k = 1}^K {\frac{1}{{L_{G,k}^{\beta}}}}.
\end{equation}

By substituting ${P_{S,B}} = \frac{{{N_B}}}{{{N_B} + K}}$ and ${P_{S,G}} = \frac{1}{{{N_B} + K}}$ into \eqref{eq:standalone_channel_power_Gain_temp}, the average channel gain under the SD scheme can finally be obtained as 
\begin{equation}
E\left\{ {{{\left| {{h_{S}}} \right|}^2}} \right\} = \frac{{\eta N_B^2}}{{L_B^{\alpha }\left( {{N_B} + K} \right)}} + \frac{{\eta {N_G}}}{{\left( {{N_B} + K} \right)}}\sum\limits_{k = 1}^K {\frac{1}{{L_{G,k}^{\beta}}}}.
\end{equation}

\section{Proof of {\textbf{Proposition \ref{prop:scd-pa-channel}}} }
\label{proof-prop:scd-pa-channel}
Let the downlink channel gain of waveguide RF port $k$ be denoted as ${p_k} = \frac{{\eta {N_G}}}{{L_{G,k}^{\beta}}}$. Assume the total transmit power is $1$, with the power allocation ratio for each port being ${w_k}$ and $\sum\nolimits_{k = 1}^K {{w_1} = 1}$. According to Equation \eqref{eq:waveguide_phase_align}, the transmitted signals from all waveguides can be coherently combined at the receiver, resulting in an equivalent channel gain of ${\left( {\sum\nolimits_{k = 1}^K {\sqrt {{p_i}{w_i}} } } \right)^2}$. We then formulate the following optimization problem
\begin{equation}
\begin{gathered}
\mathop {\arg \max }\limits_{{w_k}} {\left( {\sum\limits_{k = 1}^K {\sqrt {{p_k}{w_k}} } } \right)^2} \hfill \\
s.t.\;\sum\limits_{k = 1}^K {{w_k} = 1} ,{w_k} \geqslant 0 \hfill,
\end{gathered} 
\end{equation}
which is a convex optimization problem. Using the method of Lagrange multipliers, we can conclude that the maximum channel gain $\sum\nolimits_{k = 1}^K {{p_k}} $ is achieved when the power allocation coefficients satisfy 
\begin{equation}
w_k^* = \frac{{{p_k}}}{{\sum\nolimits_{k = 1}^K {{p_k}} }} = \frac{1}{{\sum\nolimits_{k = 1}^K {\frac{{\eta {N_G}}}{{L_{G,k}^{\beta}}}} }}\frac{{\eta {N_G}}}{{L_{G,k}^{\beta}}}.
\end{equation}
 
\section{Proof of \textbf{Proposition \ref{prop:svdscheme}}}
\label{proof-prop:svdscheme}
The derivation of \eqref{eq:semi_cooperative_channel_power_Gain} is given as follows.
{\small
\begin{equation}
\begin{gathered}
E\left\{ {{{\left| {{h_C}} \right|}^2}} \right\} \hfill \\
= E\left\{ {{{\left| {\sqrt {\frac{\eta {P_{C,B}}}{L_B^\alpha}} \left\| {{{{\bm{\tilde h}}}_B}} \right\| + \sqrt {\eta {P_{C,G}}\sum\limits_{k = 1}^K {\frac{{{N_G}}}{{L_{G,k}^{\beta}}}} } {e^{ - j{\phi _1}}}} \right|}^2}} \right\} \hfill \\
= \frac{{\eta {P_{C,B}}}}{{L_B^{\alpha }}}E\left\{ {{{\left\| {{{{\bm{\tilde h}}}_B}} \right\|}^2}} \right\} \hfill \\
+ E\left\{ {\frac{\eta }{{\sqrt{ L_B^\alpha} }}\sqrt {{P_{C,B}}{P_{C,G}}\sum\limits_{k = 1}^K {\frac{{{N_G}}}{{L_{G,k}^{\beta}}}} } \left\| {{{{\bm{\tilde h}}}_B}} \right\|\cos \left( {{\phi _1}} \right)} \right\} \hfill\\
+ \eta {P_{C,G}}\sum\limits_{k = 1}^K {\frac{{{N_G}}}{{L_{G,k}^{\beta}}}}  \hfill \\
= \frac{{\eta {N_B}{P_{C,B}}}}{{L_B^{\alpha }}} + \eta {N_G}{P_{C,G}}\sum\limits_{k = 1}^K {\frac{1}{{L_{G,k}^{\beta}}}}  \hfill 
\end{gathered} 
\end{equation}
}
The last equality holds because $\phi_1 \sim U\left(0, 2\pi \right) $ and the cross-product term is zero. With ${P_{S,B}} = \frac{{{N_B}}}{{{N_B} + K}}$ and ${P_{S,G}} = \frac{1}{{{N_B} + K}}$, the average channel gain under the SD scheme is obtained as 
\begin{equation}
E\left\{ {{{\left| {{h_C}} \right|}^2}} \right\} = \frac{{\eta N_B^2}}{{\left( {{N_B} + K} \right)L_B^{\alpha }}} + \frac{{\eta {N_G}K}}{{\left( {{N_B} + K} \right)}}\sum\limits_{k = 1}^K {\frac{1}{{L_{G,k}^{\beta}}}}.
\end{equation}

{\small
\bibliography{bibliography}

\bibliographystyle{IEEEtran}

}

\end{document}